\documentclass[prl,twocolumn,floatfix,superscriptaddress,amsmath,citeautoscript,aps,longbibliography]{revtex4-2}
\pdfoutput=1
\usepackage[T1]{fontenc}\usepackage[latin1]{inputenc}
\usepackage{dcolumn,bm,graphicx,color,booktabs,microtype,afterpage} \graphicspath{{Figures_new/}}
\usepackage[charter,greekuppercase=italicized]{mathdesign}
\renewcommand{\figurename}{Fig.}
\renewcommand{\tablename}{Table}
\makeatletter\renewcommand{\fnum@figure}[1]{\figurename~\thefigure.}\makeatother
\makeatletter\renewcommand{\fnum@table}[1]{\tablename~\thetable.}\makeatother
\newcount\hh \newcount\mm
\hh=\time \divide\hh by 60
\mm=\hh \multiply\mm by 60 \mm=-\mm
\advance\mm by \time
\def\now{\number\hh:\ifnum\mm<10{}0\fi\number\mm}
\usepackage[colorlinks,plainpages=false,linkcolor=blue,urlcolor=blue,citecolor=blue,pdfpagemode=UseNone,pdfstartview=FitBH]{hyperref}

\newcommand{\agcr}{AgCrSe$_2$}

\usepackage{soul}

\begin{document}

\title{Observation of the spiral spin liquid in a triangular-lattice material}

\author{N.~D.~Andriushin}
\affiliation{Institut f{\"u}r Festk{\"o}rper- und Materialphysik, Technische Universit{\"a}t Dresden, D-01069 Dresden, Germany}
\author{S.~E.~Nikitin}
\affiliation{Laboratory for Neutron Scattering and Imaging, PSI Center for Neutron and Muon Sciences, Paul Scherrer Institut, CH-5232 Villigen-PSI, Switzerland}
\author{{\O}.~S.~Fjellv{\aa}g}
\affiliation{Laboratory for Neutron Scattering and Imaging, PSI Center for Neutron and Muon Sciences, Paul Scherrer Institut, CH-5232 Villigen-PSI, Switzerland}
\affiliation{Department for Hydrogen Technology, Institute for Energy Technology, Kjeller, NO-2027, Norway}
\author{J.~S.~White}
\affiliation{Laboratory for Neutron Scattering and Imaging, PSI Center for Neutron and Muon Sciences, Paul Scherrer Institut, CH-5232 Villigen-PSI, Switzerland}
\author{A.~Podlesnyak}
\affiliation{Neutron Scattering Division, Oak Ridge National Laboratory, Oak Ridge, Tennessee 37831, USA}
\author{D.~S.~Inosov}
\affiliation{Institut f{\"u}r Festk{\"o}rper- und Materialphysik, Technische Universit{\"a}t Dresden, D-01069 Dresden, Germany}
\affiliation{W\"urzburg-Dresden Cluster of Excellence on Complexity and Topology in Quantum Matter\,---\,\textit{ct.qmat}, TU Dresden}
\author{M. C. Rahn}
\affiliation{Experimental Physics VI, Center for Electronic Correlations and Magnetism, University of Augsburg, 86159 Augsburg, Germany}
\affiliation{Institut f{\"u}r Festk{\"o}rper- und Materialphysik, Technische Universit{\"a}t Dresden, D-01069 Dresden, Germany}
\author{M.~Schmidt}
\affiliation{Max Planck Institute for Chemical Physics of Solids, D-01187 Dresden, Germany}
\author{M.~Baenitz}\thanks{Michael.Baenitz@cpfs.mpg.de}
\affiliation{Max Planck Institute for Chemical Physics of Solids, D-01187 Dresden, Germany}
\author{A.~S.~Sukhanov}\thanks{aleksandr.sukhanov@tu-dresden.de}
\affiliation{Experimental Physics VI, Center for Electronic Correlations and Magnetism, University of Augsburg, 86159 Augsburg, Germany}
\affiliation{Institut f{\"u}r Festk{\"o}rper- und Materialphysik, Technische Universit{\"a}t Dresden, D-01069 Dresden, Germany}

\begin{abstract}\noindent
The spiral spin liquid (SSL) is a highly degenerate state characterized by a continuous contour or surface in reciprocal space spanned by a spiral propagation vector. Although the SSL state has been predicted in a number of various theoretical models, very few materials are so far experimentally identified to host such a state. Via combined single-crystal wide-angle and small-angle neutron scattering, we report observation of the SSL in the quasi-two-dimensional delafossite-like AgCrSe$_2$. We show that it is a very close realization of the ideal Heisenberg $J_1$--$J_2$--$J_3$ frustrated model on the triangular lattice. By supplementing our experimental results with microscopic spin-dynamics simulations, we demonstrate how such exotic magnetic states are driven by thermal fluctuations and exchange frustration.

\end{abstract}

\maketitle

The phase transition on cooling from a paramagnetic to a magnetically ordered state necessarily breaks certain symmetries of the system. The static nature of magnetic order implies breaking of the time-reversal symmetry, and most often some of the rotational and/or translational symmetries are also lost. Such phase transitions take place in the vast majority of magnetic materials with non-negligible interatomic magnetic interactions. Defects and disorder-free systems of interacting spins that do not exhibit long-range order down to zero temperature are called spin liquids, which seem to be exceptionally rare in nature~\cite{Savary2017, broholm2020quantum, Yan_2024}. Spin liquids are recognized to demonstrate unconventional behavior and exotic quasi-particle excitations~\cite{shen2016evidence, Yu2018, knolle2019field} stemming from their extensive degeneracy of the ground state. The magnetic order in spin liquids is precluded by strong frustration. Instead, they feature a strongly-correlated state that still preserves the spin rotational symmetries of the paramagnetic system down to zero temperature. The spin-liquid state is often realized on a phase boundary between two ordered states, described in terms of propagation vectors anchored to the high-symmetry points of the Brillouin zone (BZ) boundary. For instance, the theoretical studies of the quantum $J_1$--$J_2$ Heisenberg model on the triangular lattice~\cite{Kaneko2014, Iqbal2016,sherman2023spectral}  revealed that the quantum spin liquid appears for a narrow parameter range between the 120$^{\circ}$ and stripe orders. However, recent works also showed that a highly-fluctuating state, akin to the quantum spin liquid, can also be formed in the classic regime at an arbitrary non-zero wavenumber. The latter was termed the spiral spin liquid (SSL) as the spins in such a state maintain a well-defined spiral pitch~\cite{Balents_2010, Wen_2019,Niggemann_2019, Bergman_2007, Yao_2021, Yan_2022, Gonzalez_2024}.

Hence, the SSL is a classical spin liquid state, which lacks a long-range order but exhibits strong short-range correlations that retain the periodicity of the spin spiral. Unlike an ordinary spin spiral state, the propagation vector of the SSL spans a contour or a surface in reciprocal space, meaning that the magnetic structure is not characterized by a singled-out propagation vector but instead has a manifold of propagation vectors that define the continuous degeneracy of the ground-state. Spins in the SSL state exhibit collective fluctuations, so the SSL-hosting materials were proposed as a promising platform for the experimental realization of emergent excitations such as subdimentional fractons, which have restricted mobility and are associated with gauge fields~\cite{Placke_2024, liu2022stacking, Yao_2021, Yan_2022, Nandkishore_2019, Pretko_2017}. Materials that were found to exhibit the SSL signatures so far are the honeycomb compound FeCl$_3$~\cite{Gao_2022}, the pyrochlore ZnCr$_2$Se$_2$~\cite{Gao_20222, Tymoshenko_2017, Inosov_2020}, the breathing-kagome lattice crystal Ca$_{10}$Cr$_7$O$_{28}$~\cite{Pohle_2021}, and the diamond-lattice material MnSc$_2$S$_4$~\cite{Gao_2016, Gao_2020}. In addition, the compound LiYbO$_2$ with an elongated diamond lattice was proposed as an SSL based on the powder-averaged data~\cite{Graham_2023}. The triangular lattice is, in turn, the prototype of a geometrically frustrated lattice, which was also predicted to host the SSL state~\cite{Mohylna_2022, Glittum_2021} yet, no experimental realizations of the SSL on a triangular lattice were reported so far.

\begin{figure*}[t]
\includegraphics[width=0.99\linewidth]{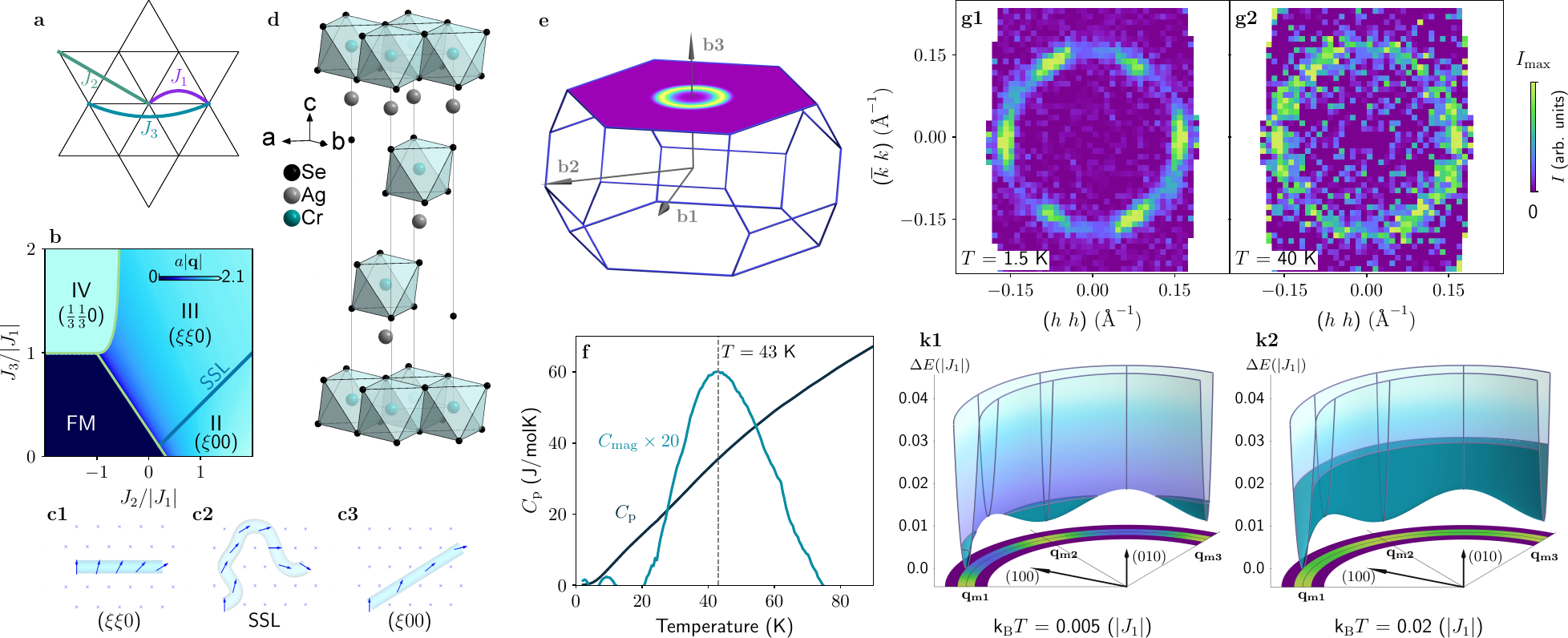}\vspace{3pt}
        \caption{(a)~Scheme of intralayer exchange interactions up to third-nearest neighbor. 
        (b)~Classical zero-temperature phase diagram in coordinates $J_2/|J_1|$ and $J_3/|J_1|$, the color represents the magnitude of the propagation vector~\cite{Glittum_2021}. 
        (c1--c3)~Schematic illustration of the cycloidal magnetic order in phase III (c1), phase II (c3) and the SSL state at the boundary between II and III (c2).
        (d,e)~Crystal structure (d) and Brillouin zone (e) of \agcr. 
        (f)~Temperature dependency of heat capacity of \agcr~\cite{Baenitz_2021}.
        (g1,g2)~The SANS-I data measured in \agcr\ at two temperatures. Note that the measured data include $\mathbf{q}$-points with $h > -0.04$, which were symmetrized to visually illustrate the angular intensity distribution.
        (k1,k2)~The classical energy calculated for exchange parameters corresponding to $\mathbf{q}_{\rm m} = (0.2\,0.2\,0)$ r.l.u. and $J_2/J_3 = 1.7$ as a function of the propagation vector of the spin cycloid. The filled part of the surface shows the level of thermal energy $k_{\rm B}T$. The colormap at the bottom shows the emergence probability of the spiral state with associated propagation vector in a form of $\exp{(-\Delta E / k_{\rm B}T)}$.
        }\label{Fig1}
\end{figure*}

In Fig.~\ref{Fig1}(b) we show the magnetic phase diagram of the triangular lattice with the $J_1\text{--}J_2\text{--}J_3$ Heisenberg exchange interactions [Fig.~\ref{Fig1}(a)] with a ferromagnetic (FM) $J_1$ (negative) and antiferromagnetic (AFM) $J_2$ and $J_3$ (both positive).
To improve the visual clarity, the phase diagram is decorated by the color-coded magnitude of the propagation vector $\bf q$, which depends on the relative ratio of exchange interactions, $J_2/J_1$ and $J_3/J_1$ (see Section~S2 in Supplemental Materials~\cite{SI}). 
In addition to the trivial FM order and commensurate $(\frac{1}{3}\,\frac{1}{3}\,0)$ AFM states, two spiral ground states can be realized: the phase II, where the propagation vector is aligned in reciprocal space as $\mathbf{q} = (\xi\,0\,0)$ [Fig.~\ref{Fig1}(c3)], and the phase III, for which the spiral propagates along an alternative direction $\mathbf{q} = (\xi\,\xi\,0)$ [Fig.~\ref{Fig1}(c1)]. In other words, the phases II and III describe the same spin spiral but oriented along each of the two principal crystallographic directions. 
The most interesting part is the boundary between the two spiral phases defined by $J_2/J_3 = 2$, for which the two orientations of propagation wavevector become degenerate.
At the critical line, the spiral propagation vector is no longer bound to the underlying lattice, and thus can arbitrarily rotate (but keeping its magnitude) on the two-dimensional plane~[see the cartoon in Fig.~\ref{Fig1}(c2)]. This leads to the characteristic singularities in its structure factor directly probed by neutron scattering, namely, to the emergence of a continuous ring of scattering intensity in reciprocal space with the radius set by the same ratio of the exchange parameters $J_2/J_3$.

Here we demonstrate that \agcr, a layered compound with the crystal structure similar to the delafossite family~\cite{li2018liquid,ding2020anharmonic,Van_Der_Lee_1989, Baenitz_2021, Han_2022, Kim_2023}, is a perfect realization of the triangular lattice with the nearest-neighbor FM exchange and the next-nearest-neighbor AFM exchange interactions. The material has a trigonal crystal structure, where the magnetic Cr$^{3+}$ ions ($S$~=~3/2) form triangular layers stacked in the ABC-fashion [Fig.\ref{Fig1}(d)]. Unlike the delafossites, the Ag ions in \agcr\ occupy only one of two triangular sublattices, leading to a non-centrosymmetric $R3m$ structure. Although the layers are weakly coupled by an AFM interaction along the trigonal $c$ axis, the physics of each triangular layer governed by the phase diagram of Fig.~\ref{Fig1}(b) remains intact. 

\begin{figure*}[t]
\includegraphics[width=0.99\linewidth]{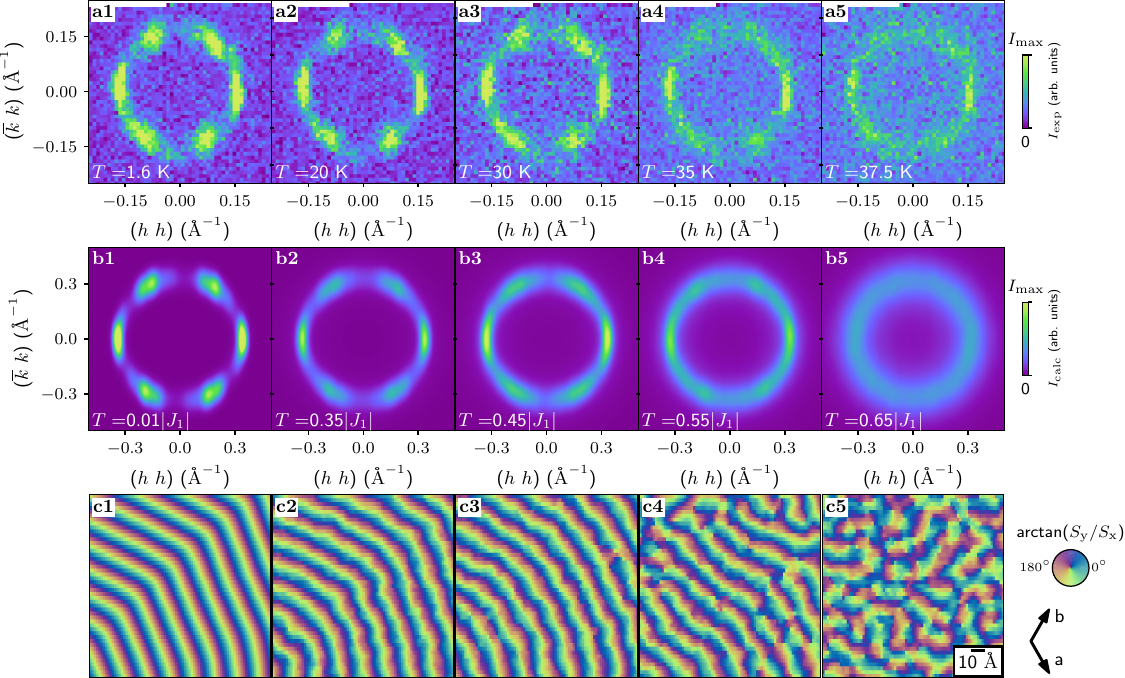}\vspace{3pt}
        \caption{(a1--a5)~The diffraction maps measured using DMC instrument at different temperatures, as indicated in each panel.
        (b1--b5)~The calculated structure factor.
        (c1--c5)~The real space spin configurations corresponding to the structure factors in (b1--b5) panels. The color represents the in-plane angle of the spins.} 
        \label{Fig2} 
\end{figure*}

The bulk properties of \agcr~are comprehensively discussed in a previous study~\cite{Baenitz_2021}. While a broad maximum was observed in the magnetic susceptibility measurements, no signatures of any phase transition could be seen in the specific heat. Instead, the extracted magnetic contribution to the total specific heat is a very broad hump with a maximum centred at $T_{\text{X}} = 43$~K, suggesting a build-up of strong magnetic fluctuations without long-range order~\cite{Baenitz_2021}. On the other hand, powder neutron diffraction at 1.5~K showed clear magnetic Bragg peaks that are sharp in the $2\theta$ angle, implying existence of magnetic correlations with a well-defined period. The Bragg peaks were indexed by an incommensurate propagation vector $\mathbf{Q} = (0.037\,0.037\,3/2)$~\cite{Baenitz_2021} corresponding to the spiral spin correlations akin to the phase III of Fig.~\ref{Fig1}(b).

In this work, we resolved the existing controversies by employing single-crystal elastic neutron scattering, in which the magnetic intensities can by fully mapped out in 3D momentum space. We show that the anomalies at $T_{\text{X}}$ occur at a cross-over from a weakly-correlated paramagnetic state to the SSL. We further corroborate our experimental observations by spin-dynamics simulations that reproduce the SSL state and its neutron scattering intensity distribution in reciprocal space, including its behavior upon temperature variation and applied magnetic field.

\section{Results}
\textbf{Observation of the SSL state.} We begin presenting our single-crystal neutron diffraction results with a brief overview of the geometry in reciprocal space. Interacting Cr spins in \agcr\ reside on the vertices of the triangular lattice in the $ab$ plane. The spins in individual triangular layers align strictly antiparallel due to an additional AFM exchange along $c$, leading to an out-of-plane component to the propagation vector, i.e. $\mathbf{Q} = \mathbf{q}_{\rm m} + (00l)$, where $l = 3/2$. This shifts the magnetic reflections to the ($H\,K\,3/2$) reciprocal plane of the first Brillouin zone (BZ), as depicted schematically in Fig.~\ref{Fig1}(e).

The diffraction map collected at $T = 40$~K, just below the maximum of a broad hump in $C_{\text{mag}}$ [Fig.~\ref{Fig1}(f)], displays a continuous ring of magnetic intensity at $|\mathbf{q}_{\rm m}|$, demonstrating that the ground state of \agcr\ at high temperature is indeed highly degenerate [Fig.~\ref{Fig1}(g2)]. It is important to note that the uniform ring retains its radial sharpness, indicating that the periodicity of the spiral correlations is well-defined (the radial broadening is limited by the instrumental resolution).

Having confirmed the formation of the SSL state at elevated temperatures, we now turn to the detailed analysis of the scattering intensity upon temperature variation. A series of diffraction patterns in the same reciprocal plane were taken at several temperatures between 1.5 and 37.5~K are shown in Figs.~\ref{Fig2}(a1-a5). One can notice that the scattering intensity undergoes a smooth transformation upon lowering the sample temperature. Namely, the initially isotropic intensity redistributes into a set of six broad maxima and minima, such that the $\langle 110 \rangle$ directions become favorable for the propagation of the spiral correlations. Because the anisotropy of scattering intensity occurs gradually upon cooling, it is in full agreement with the crossover behavior (as opposed to a phase transition) seen in the specific heat measurements [Fig.~\ref{Fig1}(f)]~\cite{Baenitz_2021}. As thermal fluctuations diminish, the full SSL degeneracy is lifted, yet the system effectively lacks the long-range order even at $T \ll T_{\text{x}}$ as the spiral propagation is never locked into a singled-out crystallographic direction. We note, that a long-range order, if existed in \agcr, would manifest itself as resolution-limited Bragg peaks.

\begin{figure*}[t]
\includegraphics[width=0.99\linewidth]{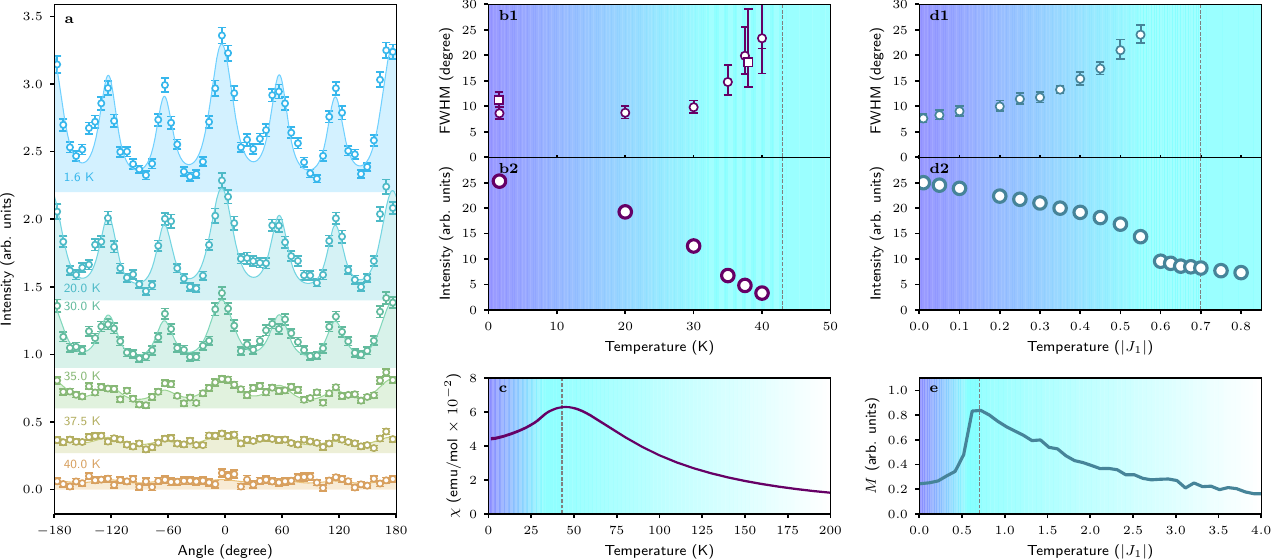}\vspace{3pt}
        \caption{(a)~The angular dependency of the intensity (DMC), the color filled peaks are the fitted Lorenzian peaks convoluted with instrumental resolution.
        (b1,b2)~The obtained fit parameters for peaks in panel (a): the Lorentzian FWHM and the intensity. The square points show the Lorentzian FWHM in SANS-I experiment, extracted similarly.
        (c)~The susceptibility data measured with in-plane field of 1~T on cooling~\cite{Baenitz_2021}.
        (d1,d2)~The calculated FWHM and the intensity.
        (e)~The magnetization calculated in field of 0.02$|J_1|$ applied along [110] direction.} \label{Fig3} 
\end{figure*}

The magnetic correlations preserve the spiral periodicity in the full temperature range below $T_{\text{x}}$, as can be seen by the resolution-limited radial width of the scattering intensity. To further quantity the spiral-orientation disorder in \agcr\ at low temperatures, we analyze its azimuthal profiles of intensity in detail. 
We described the azimuthal profiles using Lorentzian functions, whose full width at half maximum (FWHM) is proportional to the degree of disorder~\cite{osti_15009517}.

Figure~\ref{Fig3}(a) summarizes the angular profiles of intensity within the ring, which was plotted by rebinning the diffraction maps into polar coordinates. The maxima spaced by 60$^{\circ}$ were fitted with a Lorentzian function convoluted with the instrumental resolution (see Section~S1 Supplemental Materials~\cite{SI} for the details on instrumental resolution). The FWHM of the Lorentzians exhibits no change between 2 and 30~K but increases significantly at higher temperatures [Fig.\ref{Fig3}(b1)]. As was noted above, the intensity broadening takes place gradually and diverges in proximity to the crossover temperature, where separate Lorentzian profiles are no longer applicable. As can be seen, there is still a significant remnant azimuthal width of $\sim$10$^{\circ}$ (far surpassing the instrumental resolution) at 1.5 K, associated with a moderate directional disorder of the spirals even at $T \ll\ T_x$. The reliability of the extracted FWHM can be confirmed by comparing the data collected on two different instruments with distinct resolution functions (see Methods and Supplemental Materials~\cite{SI}). The extracted FWHMs from the two datasets are plotted along in Fig.\ref{Fig3}(b1)], confirming that the remnant broadening is an intrinsic property of \agcr.

It is representative to compare the crossover behavior seen in azimuthal width with the integral intensity of the whole ring. As shown in Fig.~\ref{Fig3}(b2), the observed intensity displays a gradual decrease upon warming without any sharp anomalies and does not resemble a typical critical $\sim \sqrt{T-T_\text{C}}$ behavior that is characteristic of a second-order phase transition. The absence of sharp anomalies was also evident from the magnetic susceptibility measurements in Fig.~\ref{Fig3}(c) that show a broad maximum centered at $T_{\text{x}}$~\cite{Baenitz_2021}. 

It is worth mentioning that the observed SSL state may explain the recent observation of the anomalous Hall effect in \agcr~\cite{Kim_2023}. The anomalous part of the transverse resistivity was maximized at the low temperatures but was also observed at temperatures well above $T_{\text{X}}$ without any sharp onset akin to a crossover. This suggest that the inhomogeneous spin texture of the SSL may have a nontrivial impact on the conduction electrons.

\begin{figure*}[t]
\includegraphics[width=0.99\linewidth]{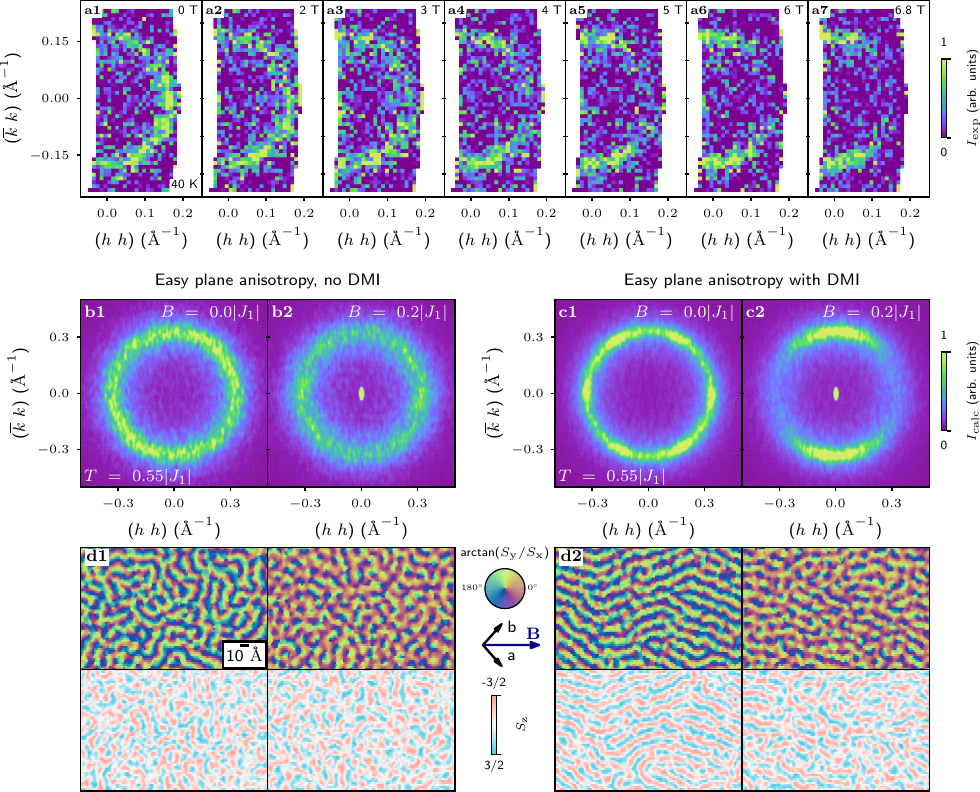}\vspace{3pt}
        \caption{(a1--a7)~The diffraction maps measured with SANS-I in different fields along [110] at 40~K.
        (b1--b2)~Comparison of simulated structure factors in absence and presence of external field applied along [11] direction (horizontal to the figure) for the model with easy plane anisotropy only.
        (c1--c2)~Simulations in the same conditions but with introduction of the DMI interaction into the model.
        (d1--d2)~The real space representation of simulated above structure factors. Each column corresponds to the (b1--c2) panel above. The top row shows the in-plane orientation of the magnetic moment as $\arctan(S_\text{y}/S_\text{x})$, the color circle in-between panels indicates the angle-color correspondence. The bottom row shows the out-of-plane component $S_\text{z}$.} \label{Fig4} 
\end{figure*}

\textbf{Modeling of the SSL state.} To simulate the SSL state in \agcr\, we considered the $J_1\text{--}J_2\text{--}J_3$ Heisenberg model with an FM $J_1$ and an AFM $J_2$ and $J_3$. We also included a weak easy-plane anisotropy $K$ that enforces coplanar spin texture with $\langle{}S_z\rangle = 0$, which was deduced in the previous study~\cite{Baenitz_2021}. To verify the adequacy of this approach, we first successfully reproduced the magnetic phase diagram of Fig.\ref{Fig2}(b1) that was originally obtained by analytic equations~\cite{Glittum_2021}. 

To reproduce the magnetic behavior of \agcr\ we chose the following model parameters: $J_2~= 0.33|J_1|$, $J_3 = 0.19|J_1|$ and $K = 0.03|J_1|$. The ground state propagation vector for these parameters equals to $(0.2\,0.2\,0)$~r.l.u., which is larger than the experimental value. Nevertheless, this parameter set allows us to capture all the general features of the SSL model at reasonable computation time. The larger propagation vector only leads to minor quantitative changes with respect to the experimental values.

The results of our simulations are presented in Figs.\ref{Fig2}(b1--b5), where the calculated structure factors are compared to the experimental patterns. For a clear comparison, we convoluted the calculated patterns with the experimental resolution. The simulations were performed at temperatures given in the units of the first exchange $J_1$, as it serves as the overall scaling of the total energy of the system. As one can see, the simulations correctly predict the anisotropic intensity distribution at low temperatures, where the azimuthal broadening shows minor variation from $T = 0.01|J_1|$ to $T = 0.45|J_1|$. The broadening is enhanced at higher $T$, and a fully isotropic ring of intensity is emerging at $T = 0.65|J_1|$ in a good agreement with the experimental pattern at 37.5~K.

We further analyzed the simulated intensities in terms of Lorentzian FWHM and plotted it in Fig.~\ref{Fig3}(d1) along with the ring integral ring intensity in Fig.~\ref{Fig3}(d2). We can conclude that the crossover behavior expected for the SSL state is fully realized in the simulations, in excellent agreement with the experimental data. Moreover, the simulated temperature dependence of magnetization, $M(T)$, closely resembles the experimental data, yet with a somewhat larger drop of the susceptibility at the lowest temperatures. The maximum of the simulated $M(T)$ is found at $T = 0.7|J_1|$. The simulations above $0.7|J_1|$ predict that the SSL state loses its radial correlations and smoothly transforms to what was previously termed a ``pancake liquid'' (see Section~S3 and S4 in Supplemental Materials~\cite{SI})~\cite{Shimokawa_2019, Yan_2022, Gonzalez_2024}. Because the intensity of the pancake liquid state is very low, it becomes indistinguishable from instrumental background in our measurements.

It is worth noting that a minor feature identified in the magnetic susceptibility in the previous work~\cite{Baenitz_2021}, namely a maximum in its temperature derivative for the in-plane fields at $\sim 32$~K corresponds to an accelerated increase of the peak broadening accompanying a small drop of the intensity seen in both the experimental data in Figs.~\ref{Fig3}(b1--b2) and our simulations in Figs.~\ref{Fig3}(d1--d2) at $T \approx 0.55K|J_1|$, which further supports the chosen exchange parameters used for the modeling.

Having achieved a good agreement between the simulated and experimental intensities in reciprocal space, we can closely examine the real-space configurations of the spins that correspond to each structure factor. The magnetic texture in real space can be presented through the rotation of the classical spin in the $ab$-plane [Figs.~\ref{Fig2}(c1--c5)], as the $S_z$ component is mostly zero across the sample plane due to the easy-plane anisotropy. At the lowest temperature [Fig.\ref{Fig2}(c1)], the plane consists of two major areas of the cycloidal modulations, where the propagation vector retains its orientation over the distances exceeding 100~\AA. The modulations in the two areas are continuously merged such that the spins of the same phase form hexagonal-shaped contours across the sample, as opposed to the distinct domains with domains walls characteristic of the ordered spiral magnets. As the temperature is elevated [Figs.\ref{Fig2}(c2-c3)], the hexagonal-shaped spin stripes become more and more concentrically bent, such that the propagation vector is found in one position over the regions less than 100~\AA. The cycloid orientation starts fluctuating already on the scale of 10~\AA~at $T = 0.55|J_1|$ [Fig.\ref{Fig2}(c4)]. Finally, the propagation direction is defined only on the scale under 10~\AA~at $T = 0.65|J_1|$ [Fig.\ref{Fig2}(c5)]. This state corresponds to the fully degenerate SSL as opposed to the states with partially lifted degeneracies at low temperatures.

\textbf{Effect of external field.} The SSL state features isotropic angular distribution of the intensity, preserving the rotational symmetry in $ab$-plane. By applying external in-plane magnetic field, one might break the rotation symmetry and potentially tune the SSL properties. However, the spins in \agcr\ are confined to the $ab$ plane~\cite{Baenitz_2021}, which makes the SSL propagation vectors insensitive to the external field direction (as soon as it is applied in plane). Indeed, the effective Zeeman field averages out to zero due to full spin winding in the $ab$ plane. Contrary to this naive expectations, the SSL state in \agcr\ is found to respond by a significant redistribution of the intensity. A field applied along (110) at 40~K gradually suppresses the spiral correlations whose propagation vector is aligned with the field $\mathbf{q}_{\rm m}~\|~\mathbf{H}$, and favors the ones with the propagation vector being orthogonal to the field $\mathbf{q}_{\rm m}~\perp~\mathbf{H}$ [Fig.~\ref{Fig4}(a1-a7)]. An apparent explanation for such behavior is that the out-of-plane spin component, which was previously precluded by the easy-plane anisotropy, becomes favorable in the applied field. Consequently, the spin correlations are no longer confined in the $ab$ plane, and hence can become sensitive to the orientation of the in-plane field. This scenario is further supported by the previous magnetization measurements, where a spin-flop transition was observed at $\sim$5~T~\cite{Baenitz_2021}.

In our simulations, the frustrated $J_1$--$J_2$--$J_3$ model with an easy-plane anisotropy indeed does not show any rotation symmetry breaking in the presence of an external field [Fig.~\ref{Fig4}(b1,b2)]. The spiral correlations maintain their orientational disorder and the pitch, while the net magnetization is build because of a slight anharmonisity of the spin spiral modulation: the magnetic moments along the spiral propagation hold on for slightly longer at the ``preferable angles'', as reflected in Fig.~\ref{Fig4}(d1) by the increased amount of red regions under the applied field. Out-of-plane spin components, as the result of an interplay between the spiral correlations and an external field, can be realized via the Dzyaloshinskii-Moriya interaction (DMI), which is allowed by the $R3m$ symmetry of \agcr. The in-plane DMI vector, which couples the nearest-neighbor spins (the $J_1$ bond), is oriented perpendicular to the bond and was assumed to have the same magnitude as the easy-plane anisotropy, namely $0.03|J_1|$. In this configuration, the DMI vector favors spirals with the spin rotation plane within the $ac$-plane and the equilibrium orientation of the rotation plane is defined by the balance between the anisotropy and the DMI.

In the absence of a field, a model augmented with the DMI is characterized by the structure factor [Fig.~\ref{Fig4}(c1)] that is very similar to the one obtained in simulations with only the easy-plane anisotropy and no DMI. 
Because the DMI interaction is weak compared to the primary isotropic interactions, it does not affect the SSL behavior, nor breaks the rotational symmetry. 

However, clear alteration in the $S_\text{z}$ component of spins is evident in the real space images [Fig~\ref{Fig4}(d2)]. The DMI causes the $S_\text{z}$ component to closely follow the spin spiral propagation. This directly influences the system's response to an external in-plane field: the spiral correlations propagating along the field direction are now disfavored as compared to the rest. The structure factor in applied field [Fig~\ref{Fig4}(c2)] then well matches the experimental data [Fig~\ref{Fig4}(a)], providing a clear evidence to the presence of DMI in the system.

It is worth noting that the neutron powder diffraction measurements~\cite{Baenitz_2021} indicated that the cycloidal modulations occur within a plane that is oriented at $89(7)^\circ$ with respect to the $c$ axis~\cite{Baenitz_2021}, possibly slightly off the $ab$ plane. This sets constraints on the possible strength of the DMI. The DMI is expected to be sufficiently weak not to overcome the easy-plane anisotropy at low temperatures in zero field, while still being strong enough to affect spiral propagation vector in the applied fields, as has been experimentally observed. However, an exact quantitative analysis of the DMI remains beyond the present study.

\section{Discussion}

It is important to note that the zero-temperature phase diagram [in Fig.~\ref{Fig1}(b)] predicts the SSL ground state exclusively at $J_2/J_3 = 2$, where the maximum level of frustration is realized. However, at finite temperatures, the SSL properties are not limited to the phase boundary of the two incommensurate states and effectively appear due to thermal fluctuations. Since the effect is mainly based on the frustration of exchange interaction, it virtually exists in the entire space of parameters compatible with an incommensurate spiral order. The temperature region where it is realistically observable heavily depends on the level of frustration, namely the proximity to the II/III phase boundary. Quantitative characteristics, such as the SSL temperature range extent and how quickly this regime develops, can change with the $J_2/J_3$ ratio and the magnitude of the propagation vector (for details see the Section~S2 of Supplementary Materials~\cite{SI}). However, the general feature of the SSL state appearance through a crossover is preserved. To further illustrate this, we calculated the Heisenberg classical energy in this model as a function of $\mathbf{q}$, see Fig.~\ref{Fig1}(k1,k2). Clearly, the energy has a deep minimum at $|\mathbf{q}_{\mathrm{m}}|$, but exhibits only a small difference between $\mathbf{q}~\|~(110)$ and $\mathbf{q}~\|~(100)$, which provides an intuitive understanding for our neutron diffraction data and the spin-dynamics calculations at finite temperatures. Essentially, thermal fluctuations first destroy the long-range order by making any orientation of the spin spiral energetically equal, and only at much higher temperatures do they destroy the spiral correlations. A rough estimation of temperature smearing is $\exp{(-\Delta E / k_{\rm B}T)}$, which gives a probability to find a spiral state with $E = E_{\rm min} + \Delta E$ in a system thermalized at temperature $T$. The corresponding colormap is shown in the lower part of Fig.~\ref{Fig1}(k1,k2).

The sizable easy-plane anisotropy in \agcr\ effectively restricts its spin dimensionality at low temperatures, which makes the XY spin model applicable for the low-energy physics of this material. Its ground state has the U(1) $\times$ U(1) symmetry, where one U(1) describes the spin rotation, whereas another U(1) stems from the momentum rotation on the spiral ring. This contrasts with the O(3) $\times$ U(1) symmetry of the Heisenberg model seemingly applicable to the SSL hosts FeCl$_3$~\cite{Gao_2022} and MnSc$_2$S$_4$~\cite{Gao_2016, Gao_2020}. Because the SSL formation within the XY model has already been quite extensively discussed in the theoretical works~\cite{Niggemann_2019, Bergman_2007, Yao_2021, Yan_2022, Gonzalez_2024, Nandkishore_2019, Placke_2024, Pretko_2017, liu2022stacking}, \agcr\ becomes an attractive playground for testing those predictions. For example, Yan and Reuther~\cite{Yan_2022} presented a topological classification of the momentum defects formed in the SSL state. After a close look into the spin configurations stabilized in our simulations, we were able to conclude that the nontopological momentum vortices are indeed realized for the model parameters of \agcr (see Section~5 of Supplementary Materials \cite{SI}). However, the momentum vortices with a non-zero topological charge (as defined in ~\cite{Yan_2022}) did not appear in our simulations, which might be related to either their higher energy costs or the fact that the system is simulated on cooling from a high temperature where the XY model is no longer applicable, which may preclude the nucleation of the topological defects. Therefore, application of the emergent higher-rank gauge theory~\cite{Yan_2022} to \agcr~remains an open question worth addressing in future studies.

To summarize, we report the observation of the SSL state by neutron scattering in a material with the perfect triangular lattice, and qualitatively reproduce our observations by the Landau-Lifshitz spin-dynamics simulations. The origin of the SSL, relying on thermal fluctuations and degeneracy, could be unraveled in terms of a specific energy landscape favoring orientational disorder at finite temperatures. The SSL behavior in external magnetic field suggests the presence of non-negligible DMI in \agcr\ providing a way of tuning the spiral correlations with an external in-plane field. Our study confirms the validity of the preceding theoretical proposals, and broadens the class of potential SSL-hosting compounds to include the triangular lattice materials, these being one of the most important model systems in the frustrated magnetism.

\section{Methods}
\noindent \textbf{Sample preparation.}

A high-quality single crystal of \agcr\  with lateral dimensions of a few mm and mass $\sim\!8$~mg was grown by chemical vapor transport using chlorine as transport agent, as described in details in \cite{Baenitz_2021}. The sample composition and crystalline quality was characterized by energy- dispersive x-ray spectroscopy, wavelength-dispersive x-ray spectroscopy, x-ray powder diffraction, Laue x-ray diffraction, and differential scanning calorimetry. The bulk magnetic properties were characterized by DC and AC magnetic measurements and specific heat measurements reported in \cite{Baenitz_2021}.

\noindent \textbf{Neutron scattering experiments.}

The neutron diffraction data were collected using the cold neutron diffractometer DMC and the small angle neutron scattering instrument SANS-I, both located in Paul Scherrer Institute (PSI), Switzerland.

In the DMC experiment, we used a PG(002) monochromator to select wavelength $\lambda = 2.45$~\AA. The in-plane $\mathbf{Q}$ resolution of DMC is considerable better than the out-of-plane, which effectively elongates the Bragg peaks in the [$\overline{1}10$] direction (perpendicular to the scattering plane). The sample was placed in a standard orange cryostat and rotated around its vertical axis with a small angle step of 0.1$^{\circ}$, such that a wide reciprocal-space volume was mapped out by the detector array covering $2\theta$ of 127$^{\circ}$ in plane and $\pm 7$$^{\circ}$ out of plane. The obtained 3D dataset was then sliced in high-symmetry planes for further analysis.

The SANS-I experiment was conducted on the same sample and the same sample orientation with respect to the scattering plane. The magnetic field was applied a using a horizontal 6.8~T cryomagnet (opening angles 45$^{\circ}$), with the magnetic field being applied along the incident neutron beam [close to the ($\overline{1}10$) reciprocal direction of the sample]. The sample and magnet were rotated around the vertical axis over 25$^{\circ}$ range in 0.5$^{\circ}$ steps to continuously span the momentum range of interest [the magnetic satellites in the $(H\,K\,3/2)$ plane]. The individual diffraction patterns collected at different sample angles by 2D detector of 0.96 $\times$ 0.96 m$^2$ were combined into a 3D dataset. The 3D dataset was then sliced in the $(H\,K\,3/2)$ plane for further analysis. Due to the geometry of the SANS-I setup, the resolution ellipsoid is elongated along the $(00L)$ reciprocal direction, whereas the resolution in the ($HK0$) plane (the plane of the SSL intensity) is significantly narrower and is approximately twice better than in the DMC measurements. For all the measurements, we subtracted the pattern collected at 50~K as a background, see Section~S4 in the Supplementary Information~\cite{SI}.

\noindent \textbf{Numerical simulations.}

The classical spin-dynamics simulations were performed with the Landau-Lifshitz dynamics approach as implemented in \textsc{Su(n)ny} program package~\cite{Zhang_2021}. A single triangular lattice layer formed with up to $300\times300$ dipolar spins and periodic boundary conditions was considered. For a detailed description, see Section~S2 in the Supplementary Information~\cite{SI}.

\bibliography{bibliography}

\begin{thebibliography}{41}%
\makeatletter
\providecommand \@ifxundefined [1]{%
 \@ifx{#1\undefined}
}%
\providecommand \@ifnum [1]{%
 \ifnum #1\expandafter \@firstoftwo
 \else \expandafter \@secondoftwo
 \fi
}%
\providecommand \@ifx [1]{%
 \ifx #1\expandafter \@firstoftwo
 \else \expandafter \@secondoftwo
 \fi
}%
\providecommand \natexlab [1]{#1}%
\providecommand \enquote  [1]{``#1''}%
\providecommand \bibnamefont  [1]{#1}%
\providecommand \bibfnamefont [1]{#1}%
\providecommand \citenamefont [1]{#1}%
\providecommand \href@noop [0]{\@secondoftwo}%
\providecommand \href [0]{\begingroup \@sanitize@url \@href}%
\providecommand \@href[1]{\@@startlink{#1}\@@href}%
\providecommand \@@href[1]{\endgroup#1\@@endlink}%
\providecommand \@sanitize@url [0]{\catcode `\\12\catcode `\$12\catcode
  `\&12\catcode `\#12\catcode `\^12\catcode `\_12\catcode `\%12\relax}%
\providecommand \@@startlink[1]{}%
\providecommand \@@endlink[0]{}%
\providecommand \url  [0]{\begingroup\@sanitize@url \@url }%
\providecommand \@url [1]{\endgroup\@href {#1}{\urlprefix }}%
\providecommand \urlprefix  [0]{URL }%
\providecommand \Eprint [0]{\href }%
\providecommand \doibase [0]{https://doi.org/}%
\providecommand \selectlanguage [0]{\@gobble}%
\providecommand \bibinfo  [0]{\@secondoftwo}%
\providecommand \bibfield  [0]{\@secondoftwo}%
\providecommand \translation [1]{[#1]}%
\providecommand \BibitemOpen [0]{}%
\providecommand \bibitemStop [0]{}%
\providecommand \bibitemNoStop [0]{.\EOS\space}%
\providecommand \EOS [0]{\spacefactor3000\relax}%
\providecommand \BibitemShut  [1]{\csname bibitem#1\endcsname}%
\let\auto@bib@innerbib\@empty
\bibitem [{\citenamefont {Savary}\ and\ \citenamefont
  {Balents}(2017)}]{Savary2017}%
  \BibitemOpen
  \bibfield  {author} {\bibinfo {author} {\bibfnamefont {L.}~\bibnamefont
  {Savary}}\ and\ \bibinfo {author} {\bibfnamefont {L.}~\bibnamefont
  {Balents}},\ }\bibfield  {title} {\bibinfo {title} {Quantum spin liquids: a
  review},\ }\href {https://doi.org/10.1088/0034-4885/80/1/016502} {\bibfield
  {journal} {\bibinfo  {journal} {Rep. Prog. Phys.}\ }\textbf {\bibinfo
  {volume} {80}},\ \bibinfo {pages} {016502} (\bibinfo {year}
  {2017})}\BibitemShut {NoStop}%
\bibitem [{\citenamefont {Broholm}\ \emph {et~al.}(2020)\citenamefont
  {Broholm}, \citenamefont {Cava}, \citenamefont {Kivelson}, \citenamefont
  {Nocera}, \citenamefont {Norman},\ and\ \citenamefont
  {Senthil}}]{broholm2020quantum}%
  \BibitemOpen
  \bibfield  {author} {\bibinfo {author} {\bibfnamefont {C.}~\bibnamefont
  {Broholm}}, \bibinfo {author} {\bibfnamefont {R.~J.}\ \bibnamefont {Cava}},
  \bibinfo {author} {\bibfnamefont {S.~A.}\ \bibnamefont {Kivelson}}, \bibinfo
  {author} {\bibfnamefont {D.~G.}\ \bibnamefont {Nocera}}, \bibinfo {author}
  {\bibfnamefont {M.~R.}\ \bibnamefont {Norman}},\ and\ \bibinfo {author}
  {\bibfnamefont {T.}~\bibnamefont {Senthil}},\ }\bibfield  {title} {\bibinfo
  {title} {Quantum spin liquids},\ }\href
  {https://doi.org/10.1126/science.aay0668} {\bibfield  {journal} {\bibinfo
  {journal} {Science}\ }\textbf {\bibinfo {volume} {367}},\ \bibinfo {pages}
  {eaay0668} (\bibinfo {year} {2020})}\BibitemShut {NoStop}%
\bibitem [{\citenamefont {Yan}\ \emph {et~al.}(2024)\citenamefont {Yan},
  \citenamefont {Benton}, \citenamefont {Moessner},\ and\ \citenamefont
  {Nevidomskyy}}]{Yan_2024}%
  \BibitemOpen
  \bibfield  {author} {\bibinfo {author} {\bibfnamefont {H.}~\bibnamefont
  {Yan}}, \bibinfo {author} {\bibfnamefont {O.}~\bibnamefont {Benton}},
  \bibinfo {author} {\bibfnamefont {R.}~\bibnamefont {Moessner}},\ and\
  \bibinfo {author} {\bibfnamefont {A.~H.}\ \bibnamefont {Nevidomskyy}},\
  }\bibfield  {title} {\bibinfo {title} {{Classification of classical spin
  liquids: Typology and resulting landscape}},\ }\href
  {https://doi.org/10.1103/PhysRevB.110.L020402} {\bibfield  {journal}
  {\bibinfo  {journal} {Phys. Rev. B}\ }\textbf {\bibinfo {volume} {110}},\
  \bibinfo {pages} {l020402} (\bibinfo {year} {2024})}\BibitemShut {NoStop}%
\bibitem [{\citenamefont {Shen}\ \emph {et~al.}(2016)\citenamefont {Shen},
  \citenamefont {Li}, \citenamefont {Wo}, \citenamefont {Li}, \citenamefont
  {Shen}, \citenamefont {Pan}, \citenamefont {Wang}, \citenamefont {Walker},
  \citenamefont {Steffens}, \citenamefont {Boehm}, \citenamefont {Hao},
  \citenamefont {Quintero-Castro}, \citenamefont {Harriger}, \citenamefont
  {Frontzek}, \citenamefont {L.}, \citenamefont {Meng}, \citenamefont {Zhang},
  \citenamefont {Chen},\ and\ \citenamefont {Zhao}}]{shen2016evidence}%
  \BibitemOpen
  \bibfield  {author} {\bibinfo {author} {\bibfnamefont {Y.}~\bibnamefont
  {Shen}}, \bibinfo {author} {\bibfnamefont {Y.-D.}\ \bibnamefont {Li}},
  \bibinfo {author} {\bibfnamefont {H.}~\bibnamefont {Wo}}, \bibinfo {author}
  {\bibfnamefont {Y.}~\bibnamefont {Li}}, \bibinfo {author} {\bibfnamefont
  {S.}~\bibnamefont {Shen}}, \bibinfo {author} {\bibfnamefont {B.}~\bibnamefont
  {Pan}}, \bibinfo {author} {\bibfnamefont {Q.}~\bibnamefont {Wang}}, \bibinfo
  {author} {\bibfnamefont {H.~C.}\ \bibnamefont {Walker}}, \bibinfo {author}
  {\bibfnamefont {P.}~\bibnamefont {Steffens}}, \bibinfo {author}
  {\bibfnamefont {M.}~\bibnamefont {Boehm}}, \bibinfo {author} {\bibfnamefont
  {Y.}~\bibnamefont {Hao}}, \bibinfo {author} {\bibfnamefont {D.~L.}\
  \bibnamefont {Quintero-Castro}}, \bibinfo {author} {\bibfnamefont {L.~W.}\
  \bibnamefont {Harriger}}, \bibinfo {author} {\bibfnamefont {M.~D.}\
  \bibnamefont {Frontzek}}, \bibinfo {author} {\bibfnamefont {H.}~\bibnamefont
  {L.}}, \bibinfo {author} {\bibfnamefont {S.}~\bibnamefont {Meng}}, \bibinfo
  {author} {\bibfnamefont {Q.}~\bibnamefont {Zhang}}, \bibinfo {author}
  {\bibfnamefont {G.}~\bibnamefont {Chen}},\ and\ \bibinfo {author}
  {\bibfnamefont {J.}~\bibnamefont {Zhao}},\ }\bibfield  {title} {\bibinfo
  {title} {{Evidence for a spinon Fermi surface in a triangular-lattice
  quantum-spin-liquid candidate}},\ }\href
  {https://doi.org/https://doi.org/10.1038/nature20614} {\bibfield  {journal}
  {\bibinfo  {journal} {Nature}\ }\textbf {\bibinfo {volume} {540}},\ \bibinfo
  {pages} {559} (\bibinfo {year} {2016})}\BibitemShut {NoStop}%
\bibitem [{\citenamefont {Yu}\ \emph {et~al.}(2018)\citenamefont {Yu},
  \citenamefont {Wang}, \citenamefont {Dong}, \citenamefont {Yao},\ and\
  \citenamefont {Li}}]{Yu2018}%
  \BibitemOpen
  \bibfield  {author} {\bibinfo {author} {\bibfnamefont {S.-L.}\ \bibnamefont
  {Yu}}, \bibinfo {author} {\bibfnamefont {W.}~\bibnamefont {Wang}}, \bibinfo
  {author} {\bibfnamefont {Z.-Y.}\ \bibnamefont {Dong}}, \bibinfo {author}
  {\bibfnamefont {Z.-J.}\ \bibnamefont {Yao}},\ and\ \bibinfo {author}
  {\bibfnamefont {J.-X.}\ \bibnamefont {Li}},\ }\bibfield  {title} {\bibinfo
  {title} {{Deconfinement of spinons in frustrated spin systems: Spectral
  perspective}},\ }\href {https://doi.org/10.1103/PhysRevB.98.134410}
  {\bibfield  {journal} {\bibinfo  {journal} {Phys. Rev. B}\ }\textbf {\bibinfo
  {volume} {98}},\ \bibinfo {pages} {134410} (\bibinfo {year}
  {2018})}\BibitemShut {NoStop}%
\bibitem [{\citenamefont {Knolle}\ and\ \citenamefont
  {Moessner}(2019)}]{knolle2019field}%
  \BibitemOpen
  \bibfield  {author} {\bibinfo {author} {\bibfnamefont {J.}~\bibnamefont
  {Knolle}}\ and\ \bibinfo {author} {\bibfnamefont {R.}~\bibnamefont
  {Moessner}},\ }\bibfield  {title} {\bibinfo {title} {A field guide to spin
  liquids},\ }\href {https://doi.org/10.1146/annurev-conmatphys-031218-013401}
  {\bibfield  {journal} {\bibinfo  {journal} {Annu. Rev. Condens. Matter
  Phys.}\ }\textbf {\bibinfo {volume} {10}},\ \bibinfo {pages} {451} (\bibinfo
  {year} {2019})}\BibitemShut {NoStop}%
\bibitem [{\citenamefont {Kaneko}\ \emph {et~al.}(2014)\citenamefont {Kaneko},
  \citenamefont {Morita},\ and\ \citenamefont {Imada}}]{Kaneko2014}%
  \BibitemOpen
  \bibfield  {author} {\bibinfo {author} {\bibfnamefont {R.}~\bibnamefont
  {Kaneko}}, \bibinfo {author} {\bibfnamefont {S.}~\bibnamefont {Morita}},\
  and\ \bibinfo {author} {\bibfnamefont {M.}~\bibnamefont {Imada}},\ }\bibfield
   {title} {\bibinfo {title} {{Gapless Spin-Liquid Phase in an Extended Spin
  1/2 Triangular Heisenberg Model}},\ }\href
  {https://doi.org/10.7566/JPSJ.83.093707} {\bibfield  {journal} {\bibinfo
  {journal} {J. Phys. Soc. Jpn.}\ }\textbf {\bibinfo {volume} {83}},\ \bibinfo
  {pages} {093707} (\bibinfo {year} {2014})}\BibitemShut {NoStop}%
\bibitem [{\citenamefont {Iqbal}\ \emph {et~al.}(2016)\citenamefont {Iqbal},
  \citenamefont {Hu}, \citenamefont {Thomale}, \citenamefont {Poilblanc},\ and\
  \citenamefont {Becca}}]{Iqbal2016}%
  \BibitemOpen
  \bibfield  {author} {\bibinfo {author} {\bibfnamefont {Y.}~\bibnamefont
  {Iqbal}}, \bibinfo {author} {\bibfnamefont {W.-J.}\ \bibnamefont {Hu}},
  \bibinfo {author} {\bibfnamefont {R.}~\bibnamefont {Thomale}}, \bibinfo
  {author} {\bibfnamefont {D.}~\bibnamefont {Poilblanc}},\ and\ \bibinfo
  {author} {\bibfnamefont {F.}~\bibnamefont {Becca}},\ }\bibfield  {title}
  {\bibinfo {title} {{Spin liquid nature in the Heisenberg triangular
  antiferromagnet}},\ }\href {https://doi.org/10.1103/PhysRevB.93.144411}
  {\bibfield  {journal} {\bibinfo  {journal} {Phys. Rev. B}\ }\textbf {\bibinfo
  {volume} {93}},\ \bibinfo {pages} {144411} (\bibinfo {year}
  {2016})}\BibitemShut {NoStop}%
\bibitem [{\citenamefont {Sherman}\ \emph {et~al.}(2023)\citenamefont
  {Sherman}, \citenamefont {Dupont},\ and\ \citenamefont
  {Moore}}]{sherman2023spectral}%
  \BibitemOpen
  \bibfield  {author} {\bibinfo {author} {\bibfnamefont {N.~E.}\ \bibnamefont
  {Sherman}}, \bibinfo {author} {\bibfnamefont {M.}~\bibnamefont {Dupont}},\
  and\ \bibinfo {author} {\bibfnamefont {J.~E.}\ \bibnamefont {Moore}},\
  }\bibfield  {title} {\bibinfo {title} {{Spectral function of the Heisenberg
  model on the triangular lattice}},\ }\href
  {https://doi.org/10.1103/PhysRevB.107.165146} {\bibfield  {journal} {\bibinfo
   {journal} {Phys. Rev. B}\ }\textbf {\bibinfo {volume} {107}},\ \bibinfo
  {pages} {165146} (\bibinfo {year} {2023})}\BibitemShut {NoStop}%
\bibitem [{\citenamefont {Balents}(2010)}]{Balents_2010}%
  \BibitemOpen
  \bibfield  {author} {\bibinfo {author} {\bibfnamefont {L.}~\bibnamefont
  {Balents}},\ }\bibfield  {title} {\bibinfo {title} {Spin liquids in
  frustrated magnets},\ }\href {https://doi.org/10.1038/nature08917} {\bibfield
   {journal} {\bibinfo  {journal} {Nature}\ }\textbf {\bibinfo {volume}
  {464}},\ \bibinfo {pages} {199} (\bibinfo {year} {2010})}\BibitemShut
  {NoStop}%
\bibitem [{\citenamefont {Wen}\ \emph {et~al.}(2019)\citenamefont {Wen},
  \citenamefont {Yu}, \citenamefont {Li}, \citenamefont {Yu},\ and\
  \citenamefont {Li}}]{Wen_2019}%
  \BibitemOpen
  \bibfield  {author} {\bibinfo {author} {\bibfnamefont {J.}~\bibnamefont
  {Wen}}, \bibinfo {author} {\bibfnamefont {S.-L.}\ \bibnamefont {Yu}},
  \bibinfo {author} {\bibfnamefont {S.}~\bibnamefont {Li}}, \bibinfo {author}
  {\bibfnamefont {W.}~\bibnamefont {Yu}},\ and\ \bibinfo {author}
  {\bibfnamefont {J.-X.}\ \bibnamefont {Li}},\ }\bibfield  {title} {\bibinfo
  {title} {Experimental identification of quantum spin liquids},\ }\href
  {https://doi.org/10.1038/s41535-019-0151-6} {\bibfield  {journal} {\bibinfo
  {journal} {npj Quantum Mater.}\ }\textbf {\bibinfo {volume} {4}},\ \bibinfo
  {pages} {12} (\bibinfo {year} {2019})}\BibitemShut {NoStop}%
\bibitem [{\citenamefont {Niggemann}\ \emph {et~al.}(2019)\citenamefont
  {Niggemann}, \citenamefont {Hering},\ and\ \citenamefont
  {Reuther}}]{Niggemann_2019}%
  \BibitemOpen
  \bibfield  {author} {\bibinfo {author} {\bibfnamefont {N.}~\bibnamefont
  {Niggemann}}, \bibinfo {author} {\bibfnamefont {M.}~\bibnamefont {Hering}},\
  and\ \bibinfo {author} {\bibfnamefont {J.}~\bibnamefont {Reuther}},\
  }\bibfield  {title} {\bibinfo {title} {Classical spiral spin liquids as a
  possible route to quantum spin liquids},\ }\href
  {https://doi.org/10.1088/1361-648X/ab4480} {\bibfield  {journal} {\bibinfo
  {journal} {J. Condens. Matter Phys.}\ }\textbf {\bibinfo {volume} {32}},\
  \bibinfo {pages} {024001} (\bibinfo {year} {2019})}\BibitemShut {NoStop}%
\bibitem [{\citenamefont {Bergman}\ \emph {et~al.}(2007)\citenamefont
  {Bergman}, \citenamefont {Alicea}, \citenamefont {Gull}, \citenamefont
  {Trebst},\ and\ \citenamefont {Balents}}]{Bergman_2007}%
  \BibitemOpen
  \bibfield  {author} {\bibinfo {author} {\bibfnamefont {D.}~\bibnamefont
  {Bergman}}, \bibinfo {author} {\bibfnamefont {J.}~\bibnamefont {Alicea}},
  \bibinfo {author} {\bibfnamefont {E.}~\bibnamefont {Gull}}, \bibinfo {author}
  {\bibfnamefont {S.}~\bibnamefont {Trebst}},\ and\ \bibinfo {author}
  {\bibfnamefont {L.}~\bibnamefont {Balents}},\ }\bibfield  {title} {\bibinfo
  {title} {Order-by-disorder and spiral spin-liquid in frustrated
  diamond-lattice antiferromagnets},\ }\href {https://doi.org/10.1038/nphys622}
  {\bibfield  {journal} {\bibinfo  {journal} {Nature Phys.}\ }\textbf {\bibinfo
  {volume} {3}},\ \bibinfo {pages} {487} (\bibinfo {year} {2007})}\BibitemShut
  {NoStop}%
\bibitem [{\citenamefont {Yao}\ \emph {et~al.}(2021)\citenamefont {Yao},
  \citenamefont {Liu}, \citenamefont {Huang}, \citenamefont {Wang},\ and\
  \citenamefont {Chen}}]{Yao_2021}%
  \BibitemOpen
  \bibfield  {author} {\bibinfo {author} {\bibfnamefont {X.-P.}\ \bibnamefont
  {Yao}}, \bibinfo {author} {\bibfnamefont {J.~Q.}\ \bibnamefont {Liu}},
  \bibinfo {author} {\bibfnamefont {C.-J.}\ \bibnamefont {Huang}}, \bibinfo
  {author} {\bibfnamefont {X.}~\bibnamefont {Wang}},\ and\ \bibinfo {author}
  {\bibfnamefont {G.}~\bibnamefont {Chen}},\ }\bibfield  {title} {\bibinfo
  {title} {Generic spiral spin liquids},\ }\href
  {https://doi.org/10.1007/s11467-021-1074-9} {\bibfield  {journal} {\bibinfo
  {journal} {Frontiers of Physics}\ }\textbf {\bibinfo {volume} {16}},\
  \bibinfo {pages} {53303} (\bibinfo {year} {2021})}\BibitemShut {NoStop}%
\bibitem [{\citenamefont {Yan}\ and\ \citenamefont {Reuther}(2022)}]{Yan_2022}%
  \BibitemOpen
  \bibfield  {author} {\bibinfo {author} {\bibfnamefont {H.}~\bibnamefont
  {Yan}}\ and\ \bibinfo {author} {\bibfnamefont {J.}~\bibnamefont {Reuther}},\
  }\bibfield  {title} {\bibinfo {title} {Low-energy structure of spiral spin
  liquids},\ }\href {https://doi.org/10.1103/PhysRevResearch.4.023175}
  {\bibfield  {journal} {\bibinfo  {journal} {Phys. Rev. Res.}\ }\textbf
  {\bibinfo {volume} {4}},\ \bibinfo {pages} {023175} (\bibinfo {year}
  {2022})}\BibitemShut {NoStop}%
\bibitem [{\citenamefont {Gonzalez}\ \emph {et~al.}(2024)\citenamefont
  {Gonzalez}, \citenamefont {Fancelli}, \citenamefont {Yan},\ and\
  \citenamefont {Reuther}}]{Gonzalez_2024}%
  \BibitemOpen
  \bibfield  {author} {\bibinfo {author} {\bibfnamefont {M.~G.}\ \bibnamefont
  {Gonzalez}}, \bibinfo {author} {\bibfnamefont {A.}~\bibnamefont {Fancelli}},
  \bibinfo {author} {\bibfnamefont {H.}~\bibnamefont {Yan}},\ and\ \bibinfo
  {author} {\bibfnamefont {J.}~\bibnamefont {Reuther}},\ }\bibfield  {title}
  {\bibinfo {title} {{Magnetic properties of the spiral spin liquid and
  surrounding phases in the square lattice XY model}},\ }\href
  {https://doi.org/10.1103/PhysRevB.110.085106} {\bibfield  {journal} {\bibinfo
   {journal} {Phys. Rev. B}\ }\textbf {\bibinfo {volume} {110}},\ \bibinfo
  {pages} {085106} (\bibinfo {year} {2024})}\BibitemShut {NoStop}%
\bibitem [{\citenamefont {Placke}\ \emph {et~al.}(2024)\citenamefont {Placke},
  \citenamefont {Benton},\ and\ \citenamefont {Moessner}}]{Placke_2024}%
  \BibitemOpen
  \bibfield  {author} {\bibinfo {author} {\bibfnamefont {B.}~\bibnamefont
  {Placke}}, \bibinfo {author} {\bibfnamefont {O.}~\bibnamefont {Benton}},\
  and\ \bibinfo {author} {\bibfnamefont {R.}~\bibnamefont {Moessner}},\
  }\bibfield  {title} {\bibinfo {title} {Ising fracton spin liquid on the
  honeycomb lattice},\ }\href {https://doi.org/10.1103/PhysRevB.110.L020401}
  {\bibfield  {journal} {\bibinfo  {journal} {Phys. Rev. B}\ }\textbf {\bibinfo
  {volume} {110}},\ \bibinfo {pages} {l020401} (\bibinfo {year}
  {2024})}\BibitemShut {NoStop}%
\bibitem [{\citenamefont {Liu}\ \emph {et~al.}(2022)\citenamefont {Liu},
  \citenamefont {Yao},\ and\ \citenamefont {Chen}}]{liu2022stacking}%
  \BibitemOpen
  \bibfield  {author} {\bibinfo {author} {\bibfnamefont {J.}~\bibnamefont
  {Liu}}, \bibinfo {author} {\bibfnamefont {X.-P.}\ \bibnamefont {Yao}},\ and\
  \bibinfo {author} {\bibfnamefont {G.}~\bibnamefont {Chen}},\ }\bibfield
  {title} {\bibinfo {title} {Stacking-induced magnetic frustration and spiral
  spin liquid},\ }\href {https://doi.org/10.1103/PhysRevB.106.L220410}
  {\bibfield  {journal} {\bibinfo  {journal} {Phys. Rev. B}\ }\textbf {\bibinfo
  {volume} {106}},\ \bibinfo {pages} {L220410} (\bibinfo {year}
  {2022})}\BibitemShut {NoStop}%
\bibitem [{\citenamefont {Nandkishore}\ and\ \citenamefont
  {Hermele}(2019)}]{Nandkishore_2019}%
  \BibitemOpen
  \bibfield  {author} {\bibinfo {author} {\bibfnamefont {R.~M.}\ \bibnamefont
  {Nandkishore}}\ and\ \bibinfo {author} {\bibfnamefont {M.}~\bibnamefont
  {Hermele}},\ }\bibfield  {title} {\bibinfo {title} {Fractons},\ }\href
  {https://doi.org/10.1146/annurev-conmatphys-031218-013604} {\bibfield
  {journal} {\bibinfo  {journal} {Annu. Rev. Condens. Matter Phys.}\ }\textbf
  {\bibinfo {volume} {10}},\ \bibinfo {pages} {295} (\bibinfo {year}
  {2019})}\BibitemShut {NoStop}%
\bibitem [{\citenamefont {Pretko}(2017)}]{Pretko_2017}%
  \BibitemOpen
  \bibfield  {author} {\bibinfo {author} {\bibfnamefont {M.}~\bibnamefont
  {Pretko}},\ }\bibfield  {title} {\bibinfo {title} {Subdimensional particle
  structure of higher rank ${U(1)}$ spin liquids},\ }\href
  {https://doi.org/10.1103/PhysRevB.95.115139} {\bibfield  {journal} {\bibinfo
  {journal} {Phys. Rev. B}\ }\textbf {\bibinfo {volume} {95}},\ \bibinfo
  {pages} {115139} (\bibinfo {year} {2017})}\BibitemShut {NoStop}%
\bibitem [{\citenamefont {Gao}\ \emph {et~al.}(2022{\natexlab{a}})\citenamefont
  {Gao}, \citenamefont {McGuire}, \citenamefont {Liu}, \citenamefont
  {Abernathy}, \citenamefont {dela Cruz}, \citenamefont {Frontzek},
  \citenamefont {Stone},\ and\ \citenamefont {Christianson}}]{Gao_2022}%
  \BibitemOpen
  \bibfield  {author} {\bibinfo {author} {\bibfnamefont {S.}~\bibnamefont
  {Gao}}, \bibinfo {author} {\bibfnamefont {M.~A.}\ \bibnamefont {McGuire}},
  \bibinfo {author} {\bibfnamefont {Y.}~\bibnamefont {Liu}}, \bibinfo {author}
  {\bibfnamefont {D.~L.}\ \bibnamefont {Abernathy}}, \bibinfo {author}
  {\bibfnamefont {C.}~\bibnamefont {dela Cruz}}, \bibinfo {author}
  {\bibfnamefont {M.}~\bibnamefont {Frontzek}}, \bibinfo {author}
  {\bibfnamefont {M.~B.}\ \bibnamefont {Stone}},\ and\ \bibinfo {author}
  {\bibfnamefont {A.~D.}\ \bibnamefont {Christianson}},\ }\bibfield  {title}
  {\bibinfo {title} {{Spiral Spin Liquid on a Honeycomb Lattice}},\ }\href
  {https://doi.org/10.1103/PhysRevLett.128.227201} {\bibfield  {journal}
  {\bibinfo  {journal} {Phys. Rev. Lett.}\ }\textbf {\bibinfo {volume} {128}},\
  \bibinfo {pages} {227201} (\bibinfo {year} {2022}{\natexlab{a}})}\BibitemShut
  {NoStop}%
\bibitem [{\citenamefont {Gao}\ \emph {et~al.}(2022{\natexlab{b}})\citenamefont
  {Gao}, \citenamefont {Pokharel}, \citenamefont {May}, \citenamefont
  {Paddison}, \citenamefont {Pasco}, \citenamefont {Liu}, \citenamefont
  {Taddei}, \citenamefont {Calder}, \citenamefont {Mandrus}, \citenamefont
  {Stone},\ and\ \citenamefont {Christianson}}]{Gao_20222}%
  \BibitemOpen
  \bibfield  {author} {\bibinfo {author} {\bibfnamefont {S.}~\bibnamefont
  {Gao}}, \bibinfo {author} {\bibfnamefont {G.}~\bibnamefont {Pokharel}},
  \bibinfo {author} {\bibfnamefont {A.~F.}\ \bibnamefont {May}}, \bibinfo
  {author} {\bibfnamefont {J.~A.~M.}\ \bibnamefont {Paddison}}, \bibinfo
  {author} {\bibfnamefont {C.}~\bibnamefont {Pasco}}, \bibinfo {author}
  {\bibfnamefont {Y.}~\bibnamefont {Liu}}, \bibinfo {author} {\bibfnamefont
  {K.~M.}\ \bibnamefont {Taddei}}, \bibinfo {author} {\bibfnamefont
  {S.}~\bibnamefont {Calder}}, \bibinfo {author} {\bibfnamefont {D.~G.}\
  \bibnamefont {Mandrus}}, \bibinfo {author} {\bibfnamefont {M.~B.}\
  \bibnamefont {Stone}},\ and\ \bibinfo {author} {\bibfnamefont {A.~D.}\
  \bibnamefont {Christianson}},\ }\bibfield  {title} {\bibinfo {title}
  {Line-graph approach to spiral spin liquids},\ }\href
  {https://doi.org/10.1103/PhysRevLett.129.237202} {\bibfield  {journal}
  {\bibinfo  {journal} {Phys. Rev. Lett.}\ }\textbf {\bibinfo {volume} {129}},\
  \bibinfo {pages} {237202} (\bibinfo {year} {2022}{\natexlab{b}})}\BibitemShut
  {NoStop}%
\bibitem [{\citenamefont {Tymoshenko}\ \emph {et~al.}(2017)\citenamefont
  {Tymoshenko}, \citenamefont {Onykiienko}, \citenamefont {M$\"{u}$ller},
  \citenamefont {Thomale}, \citenamefont {Rachel}, \citenamefont {Cameron},
  \citenamefont {Portnichenko}, \citenamefont {Efremov}, \citenamefont
  {Tsurkan}, \citenamefont {Abernathy}, \citenamefont {Ollivier}, \citenamefont
  {Schneidewind}, \citenamefont {Piovano}, \citenamefont {Felea}, \citenamefont
  {Loidl},\ and\ \citenamefont {Inosov}}]{Tymoshenko_2017}%
  \BibitemOpen
  \bibfield  {author} {\bibinfo {author} {\bibfnamefont {Y.~V.}\ \bibnamefont
  {Tymoshenko}}, \bibinfo {author} {\bibfnamefont {Y.~A.}\ \bibnamefont
  {Onykiienko}}, \bibinfo {author} {\bibfnamefont {T.}~\bibnamefont
  {M$\"{u}$ller}}, \bibinfo {author} {\bibfnamefont {R.}~\bibnamefont
  {Thomale}}, \bibinfo {author} {\bibfnamefont {S.}~\bibnamefont {Rachel}},
  \bibinfo {author} {\bibfnamefont {A.~S.}\ \bibnamefont {Cameron}}, \bibinfo
  {author} {\bibfnamefont {P.~Y.}\ \bibnamefont {Portnichenko}}, \bibinfo
  {author} {\bibfnamefont {D.~V.}\ \bibnamefont {Efremov}}, \bibinfo {author}
  {\bibfnamefont {V.}~\bibnamefont {Tsurkan}}, \bibinfo {author} {\bibfnamefont
  {D.~L.}\ \bibnamefont {Abernathy}}, \bibinfo {author} {\bibfnamefont
  {J.}~\bibnamefont {Ollivier}}, \bibinfo {author} {\bibfnamefont
  {A.}~\bibnamefont {Schneidewind}}, \bibinfo {author} {\bibfnamefont
  {A.}~\bibnamefont {Piovano}}, \bibinfo {author} {\bibfnamefont
  {V.}~\bibnamefont {Felea}}, \bibinfo {author} {\bibfnamefont
  {A.}~\bibnamefont {Loidl}},\ and\ \bibinfo {author} {\bibfnamefont {D.~S.}\
  \bibnamefont {Inosov}},\ }\bibfield  {title} {\bibinfo {title}
  {{Pseudo-Goldstone Magnons in the Frustrated $S$ = 3/2 Heisenberg Helimagnet
  ZnCr$_2$Se$_4$ with a Pyrochlore Magnetic Sublattice}},\ }\href
  {https://doi.org/10.1103/PhysRevX.7.041049} {\bibfield  {journal} {\bibinfo
  {journal} {Phys. Rev. X}\ }\textbf {\bibinfo {volume} {7}},\ \bibinfo {pages}
  {041049} (\bibinfo {year} {2017})}\BibitemShut {NoStop}%
\bibitem [{\citenamefont {Inosov}\ \emph {et~al.}(2020)\citenamefont {Inosov},
  \citenamefont {Onykiienko}, \citenamefont {Tymoshenko}, \citenamefont
  {Akopyan}, \citenamefont {Shukla}, \citenamefont {Prasai}, \citenamefont
  {Doerr}, \citenamefont {Gorbunov}, \citenamefont {Zherlitsyn}, \citenamefont
  {Voneshen}, \citenamefont {Boehm}, \citenamefont {Tsurkan}, \citenamefont
  {Felea}, \citenamefont {Loidl},\ and\ \citenamefont {Cohn}}]{Inosov_2020}%
  \BibitemOpen
  \bibfield  {author} {\bibinfo {author} {\bibfnamefont {D.~S.}\ \bibnamefont
  {Inosov}}, \bibinfo {author} {\bibfnamefont {Y.~O.}\ \bibnamefont
  {Onykiienko}}, \bibinfo {author} {\bibfnamefont {Y.~V.}\ \bibnamefont
  {Tymoshenko}}, \bibinfo {author} {\bibfnamefont {A.}~\bibnamefont {Akopyan}},
  \bibinfo {author} {\bibfnamefont {D.}~\bibnamefont {Shukla}}, \bibinfo
  {author} {\bibfnamefont {N.}~\bibnamefont {Prasai}}, \bibinfo {author}
  {\bibfnamefont {M.}~\bibnamefont {Doerr}}, \bibinfo {author} {\bibfnamefont
  {D.}~\bibnamefont {Gorbunov}}, \bibinfo {author} {\bibfnamefont
  {S.}~\bibnamefont {Zherlitsyn}}, \bibinfo {author} {\bibfnamefont {D.~J.}\
  \bibnamefont {Voneshen}}, \bibinfo {author} {\bibfnamefont {M.}~\bibnamefont
  {Boehm}}, \bibinfo {author} {\bibfnamefont {V.}~\bibnamefont {Tsurkan}},
  \bibinfo {author} {\bibfnamefont {V.}~\bibnamefont {Felea}}, \bibinfo
  {author} {\bibfnamefont {A.}~\bibnamefont {Loidl}},\ and\ \bibinfo {author}
  {\bibfnamefont {J.~L.}\ \bibnamefont {Cohn}},\ }\bibfield  {title} {\bibinfo
  {title} {Magnetic field dependence of low-energy magnons, anisotropic heat
  conduction, and spontaneous relaxation of magnetic domains in the cubic
  helimagnet {ZnCr$_2$Se$_4$}},\ }\href
  {https://doi.org/10.1103/PhysRevB.102.184431} {\bibfield  {journal} {\bibinfo
   {journal} {Phys. Rev. B}\ }\textbf {\bibinfo {volume} {102}},\ \bibinfo
  {pages} {184431} (\bibinfo {year} {2020})}\BibitemShut {NoStop}%
\bibitem [{\citenamefont {Pohle}\ \emph {et~al.}(2021)\citenamefont {Pohle},
  \citenamefont {Yan},\ and\ \citenamefont {Shannon}}]{Pohle_2021}%
  \BibitemOpen
  \bibfield  {author} {\bibinfo {author} {\bibfnamefont {R.}~\bibnamefont
  {Pohle}}, \bibinfo {author} {\bibfnamefont {H.}~\bibnamefont {Yan}},\ and\
  \bibinfo {author} {\bibfnamefont {N.}~\bibnamefont {Shannon}},\ }\bibfield
  {title} {\bibinfo {title} {{Theory of Ca$_{10}$Cr$_7$O$_{28}$ as a bilayer
  breathing-kagome magnet: Classical thermodynamics and semiclassical
  dynamics}},\ }\href
  {https://doi.org/https://doi.org/10.1103/PhysRevB.104.024426} {\bibfield
  {journal} {\bibinfo  {journal} {Phys. Rev. B}\ }\textbf {\bibinfo {volume}
  {104}},\ \bibinfo {pages} {024426} (\bibinfo {year} {2021})}\BibitemShut
  {NoStop}%
\bibitem [{\citenamefont {Gao}\ \emph {et~al.}(2016)\citenamefont {Gao},
  \citenamefont {Zaharko}, \citenamefont {Tsurkan}, \citenamefont {Su},
  \citenamefont {White}, \citenamefont {Tucker}, \citenamefont {Roessli},
  \citenamefont {Bourdarot}, \citenamefont {Sibille}, \citenamefont
  {Chernyshov}, \citenamefont {Fennell}, \citenamefont {Loidl},\ and\
  \citenamefont {R{\"{u}}egg}}]{Gao_2016}%
  \BibitemOpen
  \bibfield  {author} {\bibinfo {author} {\bibfnamefont {S.}~\bibnamefont
  {Gao}}, \bibinfo {author} {\bibfnamefont {O.}~\bibnamefont {Zaharko}},
  \bibinfo {author} {\bibfnamefont {V.}~\bibnamefont {Tsurkan}}, \bibinfo
  {author} {\bibfnamefont {Y.}~\bibnamefont {Su}}, \bibinfo {author}
  {\bibfnamefont {J.~S.}\ \bibnamefont {White}}, \bibinfo {author}
  {\bibfnamefont {G.~S.}\ \bibnamefont {Tucker}}, \bibinfo {author}
  {\bibfnamefont {B.}~\bibnamefont {Roessli}}, \bibinfo {author} {\bibfnamefont
  {F.}~\bibnamefont {Bourdarot}}, \bibinfo {author} {\bibfnamefont
  {R.}~\bibnamefont {Sibille}}, \bibinfo {author} {\bibfnamefont
  {D.}~\bibnamefont {Chernyshov}}, \bibinfo {author} {\bibfnamefont
  {T.}~\bibnamefont {Fennell}}, \bibinfo {author} {\bibfnamefont
  {A.}~\bibnamefont {Loidl}},\ and\ \bibinfo {author} {\bibfnamefont
  {C.}~\bibnamefont {R{\"{u}}egg}},\ }\bibfield  {title} {\bibinfo {title}
  {Spiral spin-liquid and the emergence of a vortex-like state in
  {MnSc}$_2${S}$_4$},\ }\href
  {https://doi.org/https://doi.org/10.1038/nphys3914} {\bibfield  {journal}
  {\bibinfo  {journal} {Nature Phys.}\ }\textbf {\bibinfo {volume} {13}},\
  \bibinfo {pages} {157} (\bibinfo {year} {2016})}\BibitemShut {NoStop}%
\bibitem [{\citenamefont {Gao}\ \emph {et~al.}(2020)\citenamefont {Gao},
  \citenamefont {Rosales}, \citenamefont {Albarrac{\'{\i}}n}, \citenamefont
  {Tsurkan}, \citenamefont {Kaur}, \citenamefont {Fennell}, \citenamefont
  {Steffens}, \citenamefont {Boehm}, \citenamefont {{\v{C}}erm{\'{a}}k},
  \citenamefont {Schneidewind}, \citenamefont {Ressouche}, \citenamefont
  {Cabra}, \citenamefont {R{\"{u}}egg},\ and\ \citenamefont
  {Zaharko}}]{Gao_2020}%
  \BibitemOpen
  \bibfield  {author} {\bibinfo {author} {\bibfnamefont {S.}~\bibnamefont
  {Gao}}, \bibinfo {author} {\bibfnamefont {H.~D.}\ \bibnamefont {Rosales}},
  \bibinfo {author} {\bibfnamefont {F.~A.~G.}\ \bibnamefont
  {Albarrac{\'{\i}}n}}, \bibinfo {author} {\bibfnamefont {V.}~\bibnamefont
  {Tsurkan}}, \bibinfo {author} {\bibfnamefont {G.}~\bibnamefont {Kaur}},
  \bibinfo {author} {\bibfnamefont {T.}~\bibnamefont {Fennell}}, \bibinfo
  {author} {\bibfnamefont {P.}~\bibnamefont {Steffens}}, \bibinfo {author}
  {\bibfnamefont {M.}~\bibnamefont {Boehm}}, \bibinfo {author} {\bibfnamefont
  {P.}~\bibnamefont {{\v{C}}erm{\'{a}}k}}, \bibinfo {author} {\bibfnamefont
  {A.}~\bibnamefont {Schneidewind}}, \bibinfo {author} {\bibfnamefont
  {E.}~\bibnamefont {Ressouche}}, \bibinfo {author} {\bibfnamefont {D.~C.}\
  \bibnamefont {Cabra}}, \bibinfo {author} {\bibfnamefont {C.}~\bibnamefont
  {R{\"{u}}egg}},\ and\ \bibinfo {author} {\bibfnamefont {O.}~\bibnamefont
  {Zaharko}},\ }\bibfield  {title} {\bibinfo {title} {Fractional
  antiferromagnetic skyrmion lattice induced by anisotropic couplings},\ }\href
  {https://doi.org/https://doi.org/10.1038/s41586-020-2716-8} {\bibfield
  {journal} {\bibinfo  {journal} {Nature}\ }\textbf {\bibinfo {volume} {586}},\
  \bibinfo {pages} {37} (\bibinfo {year} {2020})}\BibitemShut {NoStop}%
\bibitem [{\citenamefont {Graham}\ \emph {et~al.}(2023)\citenamefont {Graham},
  \citenamefont {Qureshi}, \citenamefont {Ritter}, \citenamefont {Manuel},
  \citenamefont {Wildes},\ and\ \citenamefont {Clark}}]{Graham_2023}%
  \BibitemOpen
  \bibfield  {author} {\bibinfo {author} {\bibfnamefont {J.~N.}\ \bibnamefont
  {Graham}}, \bibinfo {author} {\bibfnamefont {N.}~\bibnamefont {Qureshi}},
  \bibinfo {author} {\bibfnamefont {C.}~\bibnamefont {Ritter}}, \bibinfo
  {author} {\bibfnamefont {P.}~\bibnamefont {Manuel}}, \bibinfo {author}
  {\bibfnamefont {A.~R.}\ \bibnamefont {Wildes}},\ and\ \bibinfo {author}
  {\bibfnamefont {L.}~\bibnamefont {Clark}},\ }\bibfield  {title} {\bibinfo
  {title} {{Experimental Evidence for the Spiral Spin Liquid in LiYbO$_2$}},\
  }\href {https://doi.org/10.1103/PhysRevLett.130.166703} {\bibfield  {journal}
  {\bibinfo  {journal} {Phys. Rev. Lett.}\ }\textbf {\bibinfo {volume} {130}},\
  \bibinfo {pages} {166703} (\bibinfo {year} {2023})}\BibitemShut {NoStop}%
\bibitem [{\citenamefont {Mohylna}\ \emph {et~al.}(2022)\citenamefont
  {Mohylna}, \citenamefont {Albarrac{\'{\i}}n}, \citenamefont
  {{\v{Z}}ukovi{\v{c}}},\ and\ \citenamefont {Rosales}}]{Mohylna_2022}%
  \BibitemOpen
  \bibfield  {author} {\bibinfo {author} {\bibfnamefont {M.}~\bibnamefont
  {Mohylna}}, \bibinfo {author} {\bibfnamefont {F.~A.~G.}\ \bibnamefont
  {Albarrac{\'{\i}}n}}, \bibinfo {author} {\bibfnamefont {M.}~\bibnamefont
  {{\v{Z}}ukovi{\v{c}}}},\ and\ \bibinfo {author} {\bibfnamefont {H.~D.}\
  \bibnamefont {Rosales}},\ }\bibfield  {title} {\bibinfo {title} {Spontaneous
  antiferromagnetic skyrmion/antiskyrmion lattice and spiral spin-liquid states
  in the frustrated triangular lattice},\ }\href
  {https://doi.org/10.1103/PhysRevB.106.224406} {\bibfield  {journal} {\bibinfo
   {journal} {Phys. Rev. B}\ }\textbf {\bibinfo {volume} {106}},\ \bibinfo
  {pages} {224406} (\bibinfo {year} {2022})}\BibitemShut {NoStop}%
\bibitem [{\citenamefont {Glittum}\ and\ \citenamefont
  {Sylju{\aa}sen}(2021)}]{Glittum_2021}%
  \BibitemOpen
  \bibfield  {author} {\bibinfo {author} {\bibfnamefont {C.}~\bibnamefont
  {Glittum}}\ and\ \bibinfo {author} {\bibfnamefont {O.~F.}\ \bibnamefont
  {Sylju{\aa}sen}},\ }\bibfield  {title} {\bibinfo {title} {Arc-shaped
  structure factor in the ${J_1}\text{-}{J_2}\text{-}{J_3}$ classical
  heisenberg model on the triangular lattice},\ }\href
  {https://doi.org/10.1103/PhysRevB.104.184427} {\bibfield  {journal} {\bibinfo
   {journal} {Phys. Rev. B}\ }\textbf {\bibinfo {volume} {104}},\ \bibinfo
  {pages} {184427} (\bibinfo {year} {2021})}\BibitemShut {NoStop}%
\bibitem [{\citenamefont {Baenitz}\ \emph {et~al.}(2021)\citenamefont
  {Baenitz}, \citenamefont {Piva}, \citenamefont {Luther}, \citenamefont
  {Sichelschmidt}, \citenamefont {Ranjith}, \citenamefont {Dawczak-Debicki},
  \citenamefont {Ajeesh}, \citenamefont {Kim}, \citenamefont {Siemann},
  \citenamefont {Bigi}, \citenamefont {Manuel}, \citenamefont {Khalyavin},
  \citenamefont {Sokolov}, \citenamefont {Mokhtari}, \citenamefont {Zhang},
  \citenamefont {Yasuoka}, \citenamefont {King}, \citenamefont {Vinai},
  \citenamefont {Polewczyk}, \citenamefont {Torelli}, \citenamefont {Wosnitza},
  \citenamefont {Burkhardt}, \citenamefont {Schmidt}, \citenamefont {Rosner},
  \citenamefont {Wirth}, \citenamefont {K{\"{u}}hne}, \citenamefont {Nicklas},\
  and\ \citenamefont {Schmidt}}]{Baenitz_2021}%
  \BibitemOpen
  \bibfield  {author} {\bibinfo {author} {\bibfnamefont {M.}~\bibnamefont
  {Baenitz}}, \bibinfo {author} {\bibfnamefont {M.~M.}\ \bibnamefont {Piva}},
  \bibinfo {author} {\bibfnamefont {S.}~\bibnamefont {Luther}}, \bibinfo
  {author} {\bibfnamefont {J.}~\bibnamefont {Sichelschmidt}}, \bibinfo {author}
  {\bibfnamefont {K.~M.}\ \bibnamefont {Ranjith}}, \bibinfo {author}
  {\bibfnamefont {H.}~\bibnamefont {Dawczak-Debicki}}, \bibinfo {author}
  {\bibfnamefont {M.~O.}\ \bibnamefont {Ajeesh}}, \bibinfo {author}
  {\bibfnamefont {S.-J.}\ \bibnamefont {Kim}}, \bibinfo {author} {\bibfnamefont
  {G.}~\bibnamefont {Siemann}}, \bibinfo {author} {\bibfnamefont
  {C.}~\bibnamefont {Bigi}}, \bibinfo {author} {\bibfnamefont {P.}~\bibnamefont
  {Manuel}}, \bibinfo {author} {\bibfnamefont {D.}~\bibnamefont {Khalyavin}},
  \bibinfo {author} {\bibfnamefont {D.~A.}\ \bibnamefont {Sokolov}}, \bibinfo
  {author} {\bibfnamefont {P.}~\bibnamefont {Mokhtari}}, \bibinfo {author}
  {\bibfnamefont {H.}~\bibnamefont {Zhang}}, \bibinfo {author} {\bibfnamefont
  {H.}~\bibnamefont {Yasuoka}}, \bibinfo {author} {\bibfnamefont {P.~D.~C.}\
  \bibnamefont {King}}, \bibinfo {author} {\bibfnamefont {G.}~\bibnamefont
  {Vinai}}, \bibinfo {author} {\bibfnamefont {V.}~\bibnamefont {Polewczyk}},
  \bibinfo {author} {\bibfnamefont {P.}~\bibnamefont {Torelli}}, \bibinfo
  {author} {\bibfnamefont {J.}~\bibnamefont {Wosnitza}}, \bibinfo {author}
  {\bibfnamefont {U.}~\bibnamefont {Burkhardt}}, \bibinfo {author}
  {\bibfnamefont {B.}~\bibnamefont {Schmidt}}, \bibinfo {author} {\bibfnamefont
  {H.}~\bibnamefont {Rosner}}, \bibinfo {author} {\bibfnamefont
  {S.}~\bibnamefont {Wirth}}, \bibinfo {author} {\bibfnamefont
  {H.}~\bibnamefont {K{\"{u}}hne}}, \bibinfo {author} {\bibfnamefont
  {M.}~\bibnamefont {Nicklas}},\ and\ \bibinfo {author} {\bibfnamefont
  {M.}~\bibnamefont {Schmidt}},\ }\bibfield  {title} {\bibinfo {title} {Planar
  triangular {A}g{C}r{S}e$_2$: Magnetic frustration, short range correlations,
  and field-tuned anisotropic cycloidal magnetic order},\ }\href
  {https://doi.org/10.1103/PhysRevB.104.134410} {\bibfield  {journal} {\bibinfo
   {journal} {Phys. Rev. B}\ }\textbf {\bibinfo {volume} {104}},\ \bibinfo
  {pages} {134410} (\bibinfo {year} {2021})}\BibitemShut {NoStop}%
\bibitem [{SI()}]{SI}%
  \BibitemOpen
  \href@noop {} {}\bibinfo {howpublished} {See Supplemental Material at the
  link, for the description of the spin dynamics simulations and the details of
  fitting of the neutron diffraction.}\BibitemShut {Stop}%
\bibitem [{\citenamefont {Li}\ \emph {et~al.}(2018)\citenamefont {Li},
  \citenamefont {Wang}, \citenamefont {Kawakita}, \citenamefont {Zhang},
  \citenamefont {Feygenson}, \citenamefont {Yu}, \citenamefont {Wu},
  \citenamefont {Ohara}, \citenamefont {Kikuchi}, \citenamefont {Shibata},
  \citenamefont {Yamada}, \citenamefont {Ning}, \citenamefont {Chen},
  \citenamefont {He}, \citenamefont {Vaknin}, \citenamefont {Wu}, \citenamefont
  {Nakajima},\ and\ \citenamefont {Kanatzidis}}]{li2018liquid}%
  \BibitemOpen
  \bibfield  {author} {\bibinfo {author} {\bibfnamefont {B.}~\bibnamefont
  {Li}}, \bibinfo {author} {\bibfnamefont {H.}~\bibnamefont {Wang}}, \bibinfo
  {author} {\bibfnamefont {Y.}~\bibnamefont {Kawakita}}, \bibinfo {author}
  {\bibfnamefont {Q.}~\bibnamefont {Zhang}}, \bibinfo {author} {\bibfnamefont
  {M.}~\bibnamefont {Feygenson}}, \bibinfo {author} {\bibfnamefont {H.~L.}\
  \bibnamefont {Yu}}, \bibinfo {author} {\bibfnamefont {D.}~\bibnamefont {Wu}},
  \bibinfo {author} {\bibfnamefont {K.}~\bibnamefont {Ohara}}, \bibinfo
  {author} {\bibfnamefont {T.}~\bibnamefont {Kikuchi}}, \bibinfo {author}
  {\bibfnamefont {K.}~\bibnamefont {Shibata}}, \bibinfo {author} {\bibfnamefont
  {T.}~\bibnamefont {Yamada}}, \bibinfo {author} {\bibfnamefont {X.~K.}\
  \bibnamefont {Ning}}, \bibinfo {author} {\bibfnamefont {Y.}~\bibnamefont
  {Chen}}, \bibinfo {author} {\bibfnamefont {J.~Q.}\ \bibnamefont {He}},
  \bibinfo {author} {\bibfnamefont {D.}~\bibnamefont {Vaknin}}, \bibinfo
  {author} {\bibfnamefont {R.~Q.}\ \bibnamefont {Wu}}, \bibinfo {author}
  {\bibfnamefont {K.}~\bibnamefont {Nakajima}},\ and\ \bibinfo {author}
  {\bibfnamefont {M.~G.}\ \bibnamefont {Kanatzidis}},\ }\bibfield  {title}
  {\bibinfo {title} {Liquid-like thermal conduction in intercalated layered
  crystalline solids},\ }\href {https://doi.org/10.1038/s41563-017-0004-2}
  {\bibfield  {journal} {\bibinfo  {journal} {Nature Mater.}\ }\textbf
  {\bibinfo {volume} {17}},\ \bibinfo {pages} {226} (\bibinfo {year}
  {2018})}\BibitemShut {NoStop}%
\bibitem [{\citenamefont {Ding}\ \emph {et~al.}(2020)\citenamefont {Ding},
  \citenamefont {Niedziela}, \citenamefont {Bansal}, \citenamefont {Wang},
  \citenamefont {He}, \citenamefont {May}, \citenamefont {Ehlers},
  \citenamefont {Abernathy}, \citenamefont {Said}, \citenamefont {Alatas},
  \citenamefont {Reng}, \citenamefont {Arya},\ and\ \citenamefont
  {Delaire}}]{ding2020anharmonic}%
  \BibitemOpen
  \bibfield  {author} {\bibinfo {author} {\bibfnamefont {J.}~\bibnamefont
  {Ding}}, \bibinfo {author} {\bibfnamefont {J.~L.}\ \bibnamefont {Niedziela}},
  \bibinfo {author} {\bibfnamefont {D.}~\bibnamefont {Bansal}}, \bibinfo
  {author} {\bibfnamefont {J.}~\bibnamefont {Wang}}, \bibinfo {author}
  {\bibfnamefont {X.}~\bibnamefont {He}}, \bibinfo {author} {\bibfnamefont
  {A.~F.}\ \bibnamefont {May}}, \bibinfo {author} {\bibfnamefont
  {G.}~\bibnamefont {Ehlers}}, \bibinfo {author} {\bibfnamefont {D.~L.}\
  \bibnamefont {Abernathy}}, \bibinfo {author} {\bibfnamefont {A.}~\bibnamefont
  {Said}}, \bibinfo {author} {\bibfnamefont {A.}~\bibnamefont {Alatas}},
  \bibinfo {author} {\bibfnamefont {Y.}~\bibnamefont {Reng}}, \bibinfo {author}
  {\bibfnamefont {G.}~\bibnamefont {Arya}},\ and\ \bibinfo {author}
  {\bibfnamefont {O.}~\bibnamefont {Delaire}},\ }\bibfield  {title} {\bibinfo
  {title} {{Anharmonic lattice dynamics and superionic transition in
  AgCrSe$_2$}},\ }\href {https://doi.org/10.1073/pnas.1913916117} {\bibfield
  {journal} {\bibinfo  {journal} {Proc. Natl. Acad. Sci. U.S.A.}\ }\textbf
  {\bibinfo {volume} {117}},\ \bibinfo {pages} {3930} (\bibinfo {year}
  {2020})}\BibitemShut {NoStop}%
\bibitem [{\citenamefont {Lee}\ and\ \citenamefont
  {Wiegers}(1989)}]{Van_Der_Lee_1989}%
  \BibitemOpen
  \bibfield  {author} {\bibinfo {author} {\bibfnamefont {A.~V.~D.}\
  \bibnamefont {Lee}}\ and\ \bibinfo {author} {\bibfnamefont {G.}~\bibnamefont
  {Wiegers}},\ }\bibfield  {title} {\bibinfo {title} {{Anharmonic thermal
  motion of Ag in AgCrSe$_2$: A high-temperature single-crystal X-ray
  diffraction study}},\ }\href {https://doi.org/10.1016/0022-4596(89)90285-5}
  {\bibfield  {journal} {\bibinfo  {journal} {J. Solid State Chem.}\ }\textbf
  {\bibinfo {volume} {82}},\ \bibinfo {pages} {216} (\bibinfo {year}
  {1989})}\BibitemShut {NoStop}%
\bibitem [{\citenamefont {Han}\ \emph {et~al.}(2022)\citenamefont {Han},
  \citenamefont {Qi}, \citenamefont {Huang}, \citenamefont {Wang},
  \citenamefont {Li},\ and\ \citenamefont {Zhang}}]{Han_2022}%
  \BibitemOpen
  \bibfield  {author} {\bibinfo {author} {\bibfnamefont {D.}~\bibnamefont
  {Han}}, \bibinfo {author} {\bibfnamefont {J.}~\bibnamefont {Qi}}, \bibinfo
  {author} {\bibfnamefont {Y.}~\bibnamefont {Huang}}, \bibinfo {author}
  {\bibfnamefont {Z.}~\bibnamefont {Wang}}, \bibinfo {author} {\bibfnamefont
  {B.}~\bibnamefont {Li}},\ and\ \bibinfo {author} {\bibfnamefont
  {Z.}~\bibnamefont {Zhang}},\ }\bibfield  {title} {\bibinfo {title}
  {{Anisotropic magnetoelectric transport in AgCrSe$_2$ single crystals}},\
  }\href {https://doi.org/10.1063/5.0120748} {\bibfield  {journal} {\bibinfo
  {journal} {Appl. Phys. Lett.}\ }\textbf {\bibinfo {volume} {121}},\ \bibinfo
  {pages} {182405} (\bibinfo {year} {2022})}\BibitemShut {NoStop}%
\bibitem [{\citenamefont {Kim}\ \emph {et~al.}(2023)\citenamefont {Kim},
  \citenamefont {Zhu}, \citenamefont {Piva}, \citenamefont {Schmidt},
  \citenamefont {Fartab}, \citenamefont {Mackenzie}, \citenamefont {Baenitz},
  \citenamefont {Nicklas}, \citenamefont {Rosner}, \citenamefont {Cook},
  \citenamefont {Gonz\'{a}lez-Hern\'{a}ndez}, \citenamefont {\u{S}mejkal},\
  and\ \citenamefont {Zhang}}]{Kim_2023}%
  \BibitemOpen
  \bibfield  {author} {\bibinfo {author} {\bibfnamefont {S.-J.}\ \bibnamefont
  {Kim}}, \bibinfo {author} {\bibfnamefont {J.}~\bibnamefont {Zhu}}, \bibinfo
  {author} {\bibfnamefont {M.~M.}\ \bibnamefont {Piva}}, \bibinfo {author}
  {\bibfnamefont {M.}~\bibnamefont {Schmidt}}, \bibinfo {author} {\bibfnamefont
  {D.}~\bibnamefont {Fartab}}, \bibinfo {author} {\bibfnamefont {A.~P.}\
  \bibnamefont {Mackenzie}}, \bibinfo {author} {\bibfnamefont {M.}~\bibnamefont
  {Baenitz}}, \bibinfo {author} {\bibfnamefont {M.}~\bibnamefont {Nicklas}},
  \bibinfo {author} {\bibfnamefont {H.}~\bibnamefont {Rosner}}, \bibinfo
  {author} {\bibfnamefont {A.~M.}\ \bibnamefont {Cook}}, \bibinfo {author}
  {\bibfnamefont {R.}~\bibnamefont {Gonz\'{a}lez-Hern\'{a}ndez}}, \bibinfo
  {author} {\bibfnamefont {L.}~\bibnamefont {\u{S}mejkal}},\ and\ \bibinfo
  {author} {\bibfnamefont {H.}~\bibnamefont {Zhang}},\ }\bibfield  {title}
  {\bibinfo {title} {Observation of the anomalous hall effect in a layered
  polar semiconductor},\ }\href {https://doi.org/10.1002/advs.202307306}
  {\bibfield  {journal} {\bibinfo  {journal} {Advanced Science}\ }\textbf
  {\bibinfo {volume} {11}},\ \bibinfo {pages} {2307306} (\bibinfo {year}
  {2023})}\BibitemShut {NoStop}%
\bibitem [{\citenamefont {Zaliznyak}\ and\ \citenamefont
  {Lee}(2004)}]{osti_15009517}%
  \BibitemOpen
  \bibfield  {author} {\bibinfo {author} {\bibfnamefont {I.~A.}\ \bibnamefont
  {Zaliznyak}}\ and\ \bibinfo {author} {\bibfnamefont {S.~H.}\ \bibnamefont
  {Lee}},\ }\bibfield  {title} {\bibinfo {title} {{Magnetic Neutron
  Scattering}}\ }\href {https://www.osti.gov/biblio/15009517} {} (\bibinfo
  {year} {2004})\BibitemShut {NoStop}%
\bibitem [{\citenamefont {Shimokawa}\ and\ \citenamefont
  {Kawamura}(2019)}]{Shimokawa_2019}%
  \BibitemOpen
  \bibfield  {author} {\bibinfo {author} {\bibfnamefont {T.}~\bibnamefont
  {Shimokawa}}\ and\ \bibinfo {author} {\bibfnamefont {H.}~\bibnamefont
  {Kawamura}},\ }\bibfield  {title} {\bibinfo {title} {Ripple state in the
  frustrated honeycomb-lattice antiferromagnet},\ }\href
  {https://doi.org/10.1103/PhysRevLett.123.057202} {\bibfield  {journal}
  {\bibinfo  {journal} {Phys. Rev. Lett.}\ }\textbf {\bibinfo {volume} {123}},\
  \bibinfo {pages} {057202} (\bibinfo {year} {2019})}\BibitemShut {NoStop}%
\bibitem [{\citenamefont {Zhang}\ and\ \citenamefont
  {Batista}(2021)}]{Zhang_2021}%
  \BibitemOpen
  \bibfield  {author} {\bibinfo {author} {\bibfnamefont {H.}~\bibnamefont
  {Zhang}}\ and\ \bibinfo {author} {\bibfnamefont {C.~D.}\ \bibnamefont
  {Batista}},\ }\bibfield  {title} {\bibinfo {title} {Classical spin dynamics
  based on {SU}(${N}$) coherent states},\ }\href
  {https://doi.org/10.1103/PhysRevB.104.104409} {\bibfield  {journal} {\bibinfo
   {journal} {Phys. Rev. B}\ }\textbf {\bibinfo {volume} {104}},\ \bibinfo
  {pages} {104409} (\bibinfo {year} {2021})}\BibitemShut {NoStop}%
\bibitem [{\citenamefont {Dahlbom}\ \emph {et~al.}(2022)\citenamefont
  {Dahlbom}, \citenamefont {Miles}, \citenamefont {Zhang}, \citenamefont
  {Batista},\ and\ \citenamefont {Barros}}]{Dahlbom_2022}%
  \BibitemOpen
  \bibfield  {author} {\bibinfo {author} {\bibfnamefont {D.}~\bibnamefont
  {Dahlbom}}, \bibinfo {author} {\bibfnamefont {C.}~\bibnamefont {Miles}},
  \bibinfo {author} {\bibfnamefont {H.}~\bibnamefont {Zhang}}, \bibinfo
  {author} {\bibfnamefont {C.~D.}\ \bibnamefont {Batista}},\ and\ \bibinfo
  {author} {\bibfnamefont {K.}~\bibnamefont {Barros}},\ }\bibfield  {title}
  {\bibinfo {title} {{Langevin dynamics of generalized spins as {SU}($N$)
  coherent states}},\ }\href {https://doi.org/10.1103/PhysRevB.106.235154}
  {\bibfield  {journal} {\bibinfo  {journal} {Phys. Rev. B}\ }\textbf {\bibinfo
  {volume} {106}},\ \bibinfo {pages} {235154} (\bibinfo {year}
  {2022})}\BibitemShut {NoStop}%
\end{thebibliography}%

\smallskip
\noindent\textbf{Data availability}
All relevant data are available from the authors upon reasonable request.

\noindent\textbf{Competing interests}
The authors declare no competing interests.

\noindent\textbf{Author Contribution Statement}
A.S.S, M.B. and S.E.N. conceived the idea and designed the experiments. Crystals were grown by M.S. Neutron scattering was performed by S.E.N., A.S.S., \O.S.F., J.S.W., A.P. and the data were analyzed by N.D.A., S.E.N. and A.S.S. Simulations and analysis were performed by N.D.A., A.S.S. and S.E.N. The manuscript was written by A.S.S., N.D.A. and S.E.N with assistance of M.C.R. and D.S.I. and all authors discussed the results and commented the manuscript.

\noindent\textbf{Acknowledgments}
We thank V.~Hasse for technical support in crystal growth. We acknowledge financial support from the Swiss National Science Foundation, from the European Research Council under the grant Hyper Quantum Criticality (HyperQC). N.D.A. and D.S.I. are greatful for support of the German Research Foundation (DFG) through the Collaborative Research Center SFB 1143 (project \# 247310070); through the W\"urzburg-Dresden Cluster of Excellence on Complexity and Topology in Quantum Materials\,---\,\textit{ct.qmat} (EXC~2147, Project No.\ 390858490). This work is based on experiments performed at the Swiss spallation neutron source SINQ, Paul Scherrer Institute, Villigen, Switzerland. M.C.R. is grateful for support through the Emmy-Noether program of the DFG (project-id 501391385). \smallskip

\vfill
\onecolumngrid\clearpage

\renewcommand\thesection{S\arabic{section}}
\renewcommand\thefigure{S\arabic{figure}}
\renewcommand\thetable{S\arabic{table}}
\renewcommand\theequation{S\arabic{equation}}

\makeatletter
\setcounter{page}{1}\setcounter{figure}{0}\setcounter{table}{0}\setcounter{equation}{0}
\setcounter{secnumdepth}{1}
\onecolumngrid\normalsize

\begin{center}{\vspace*{0.1pt}\Large{Supplementary Information \smallskip\\\sl\textbf{``\hspace{1pt}Observation of the spiral spin liquid in a triangular-lattice material''}}}\end{center}\bigskip

\vskip4mm

\centerline{N.~D.~Andriushin, S.~E.~Nikitin, O.~S.~Fjellvag, J.~S.~White, A.~Podlesnyak,} 

\centerline{D.~S.~Inosov, M.~C.~Rahn, M.~Schmidt, M.~Baenitz, and A.~S.~Sukhanov} 

\vskip10mm
\twocolumngrid

\section{Instrumental resolution in experimental data} \label{SIres}

\subsection{DMC experiment}

\begin{figure}
\includegraphics[width=0.99\linewidth]{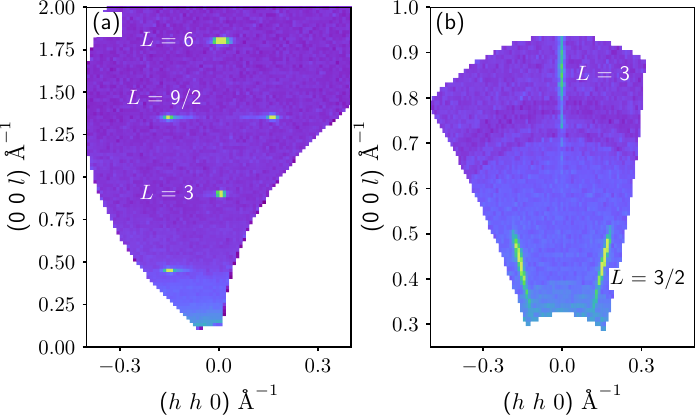}\vspace{3pt}
        \caption{~(a)~The scattering plane in the DMC experiment.
        (b)~The scattering plane in the SANS-I experiment. }
        \label{FigS1}
\end{figure}

\begin{figure}
\includegraphics[width=0.99\linewidth]{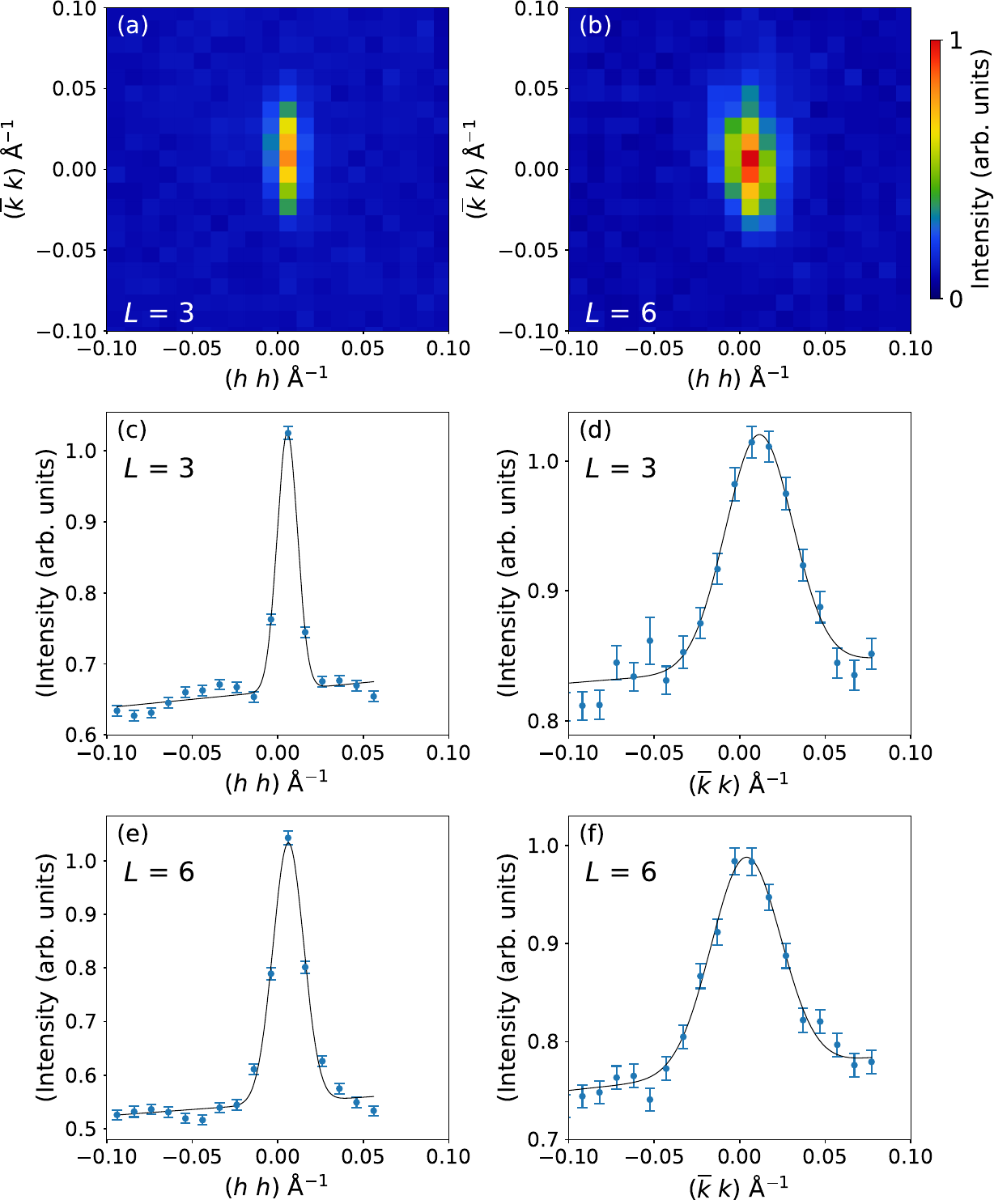}\vspace{3pt}
        \caption{~The instrumental resolution in the DMC experiment determined by the nuclear reflections (003) and (006).
        (a,b)~The intensity maps in the $(H\,K\,3)$ (a) and $(H\,K\,6)$ (b) reciprocal-space planes.
        (c,e)~The intensity profiles of the (003) and (006) peaks along the $(h\,h\,0)$ direction, respectively.
        (d,f)~The intensity profiles of the (003) and (006) peaks along the $(-k\,k\,0)$ direction, respectively.}
        \label{FigS2}
\end{figure}

\begin{figure}
\includegraphics[width=0.99\linewidth]{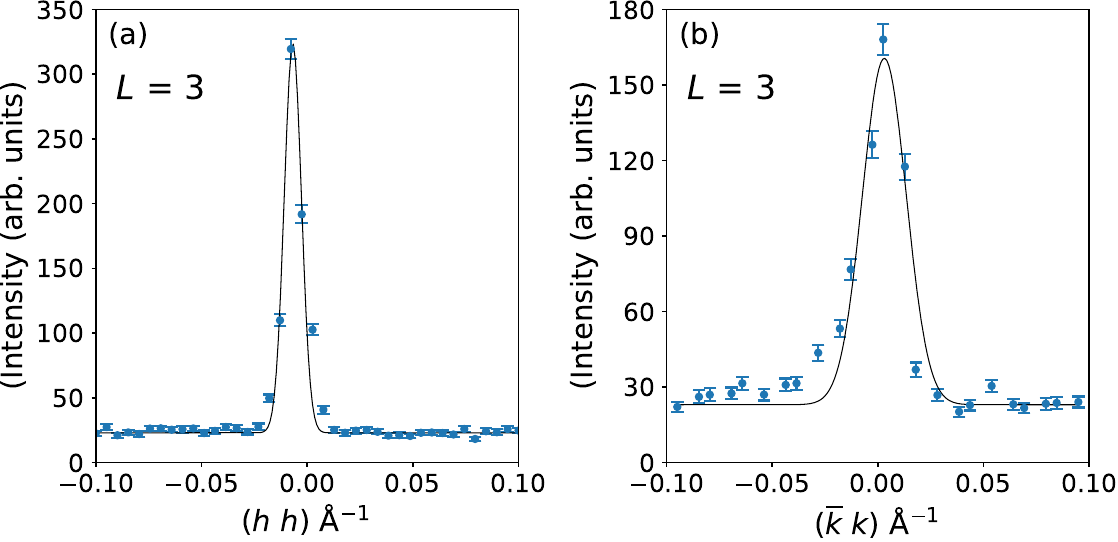}\vspace{3pt}
        \caption{~The Gaussian fits to the nuclear (003) Bragg peak in the SANS-I experiment that show the instrumental resolution for the reciprocal $(hh0)$ (a) and $(-kk0)$ directions. The symbols are data, the solid lines are the fits.}
        \label{FigS3}
\end{figure}

The SSL state in \agcr\ was confirmed by observation of a uniform ring in neutron diffraction data using the neutron diffractometer DMC and the neutron scattering instrument SANS-I. Before performing detailed analysis of the magnetic peaks, we first determined the instrumental resolution in both experimental setups, as this is required when magnetic-peak broadening is extracted. Because the resolution profile is strikingly different for DMC and SANS-I, the internal broadening can be reliably determined via cross analysis of the two datasets. 

Because of high quality of the sample, the crystal-structure (nuclear) peaks were assumed resolution limited. Therefore, fits of the nuclear peaks allowed us to determine the instrumental resolution function. The observed broadening of the magnetic Bragg peaks can therefore be associated with intrinsic properties of the magnetic system.

As the magnetic reflections of interest are located close to $(0\,0\,4.5)$ r.l.u., we used the data on the nuclear Bragg peaks nearby: $(0\,0\,3)$ and $(0\,0\,6)$. The full $(H\,K\,4.5)$ maps of these peaks are shown in Figs.~\ref{FigS2}(a,b). The peak shape forms an ellipse elongated along one of the principal crystallographic directions. Therefore, we considered two perpendicular intensity profiles and fitted them with a Gaussian function, which yields the FWHM of the peak along the reciprocal $(h\,h\,0)$ and $(-k\,k\,0)$ directions [Figs.~\ref{FigS2}(c)--(f)]. Since the resolution depends on momentum transfer, the peak at (006) has larger FWHM, as can be seen from the corresponding profiles.

The magnetic satellites $(\xi\,\xi\,4.5)$ are almost in between of these two structural reflections, meaning that the experimental resolution in the vicinity of $(0\,0\,4.5)$ can be well interpolated by an average of the two FWHM values of the $(0\,0\,3)$ and $(0\,0\,6)$ peaks. The interpolated values for the $(\xi~\xi~4.5)$ plane are then $0.018~\text{\AA}^{-1}$ along $(h\,h\,0)$ and $0.047~\text{\AA}^{-1}$ for $(-k\,k\,0)$.

For analysis of orientational disordering in \agcr, the diffraction pattern was rebinned into polar coordinates, allowing the extraction of azimuthal (in-plane angle) width to be done in a convenient way. For this purpose, the instrumental resolution can be directly recalculated as a function of azimuthal angle. For the zero azimuthal angle [the $(h\,h\,0)$ direction] and $|\mathbf{q}|~=~0.15~\text{\AA}^{-1}$, the instrumental angular FWHM $a = 16.7$~degrees. For the orthogonal direction [the $(-k\,k\,0)$ direction] it is determined to be $b = 6.5$~degrees. During the fitting, the angular resolution function at any particular angle is assumed to be a Gaussian with the FWHM taken as interpolation between these two values: $\text{FWHM} = a\cos^2{\phi} + b\sin^2{\phi}$.

\subsection{SANS-I experiment}

The momentum resolution profile of the SANS-I instrument has form distinct from that of DMC. Namely, the resolution ellipsoid is elongated radially in reciprocal space, i.e. along $(00L)$ for the nuclear (003) Bragg peak, as can be seen in the $(HHL)$ map in Fig.~\ref{FigS1}(b). The cross-section of the resolution ellipsoid in the perpendicular $(HK3)$ (crossing the nuclear peak) and $(H\,K\,3/2)$ (crossing the magnetic peaks) planes is also an ellipsoid for which the FWHM along $(h\,h\,0)$ and $(-k\,k\,0)$ directions differs by a factor of $\sim$2.5, which is similar to the DMC measurements. However, for both directions the FWHM is smaller than in the DMC setup, allowing for better refinement of the intrinsic SSL broadening. Figures~\ref{FigS3}(a) and \ref{FigS3}(b) demonstrate Gaussian fits to the (003) peak along $(h\,h\,0)$ and $(-k\,k\,0)$ directions, respectively. The resulted instrumental resolution was then found as FWHM($hh0$)~=~0.0096(3)~\AA$^{-1}$ and FWHM($-kk0$)~=~0.0244(14)~\AA$^{-1}$. The FWHM in \AA$^{-1}$ can again be recalculated into FWHM in degree for azimuthal cuts through the SSL state, which for the SANS-I measurements resulted in 1.8 and 4.6 degrees for $(h\,h\,0)$ and $(-k\,k\,0)$ directions respectively.

\section{Calculation details} \label{SIcalc}

\begin{figure}
\includegraphics[width=0.99\linewidth]{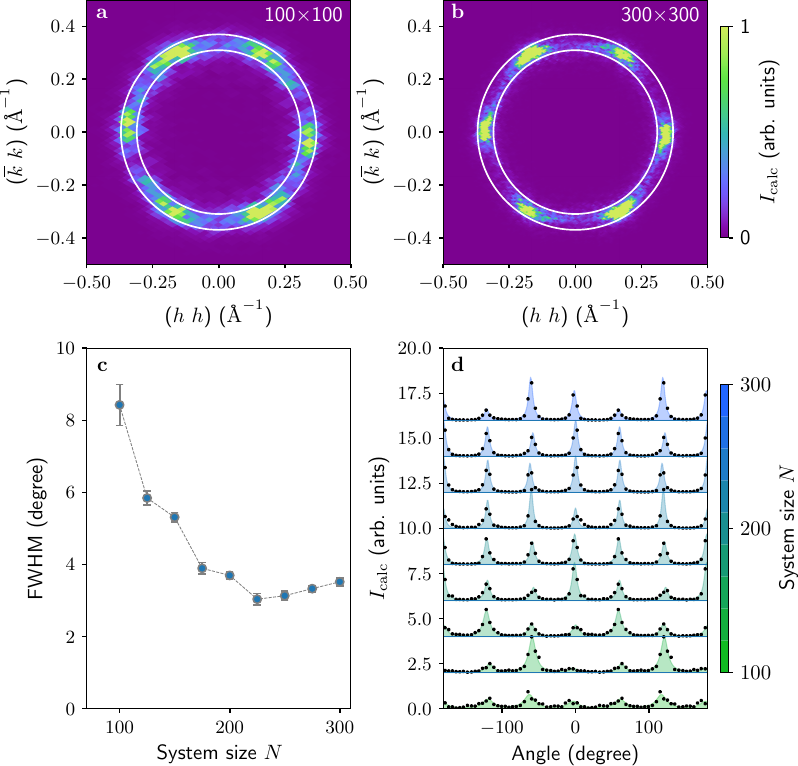}\vspace{3pt}
        \caption{~The system-size tests.
        (a,b)~Calculated structure factor for $100\times100$ and $300\times300$ systems.
        (c)~Dependence of the azimuthal width on the system size.
        (d)~Azimuthal dependence of the intensity obtained by an integration inside the area highlighted by the two white circles in (a) and (b). The color shading shows the Lorentzian fits, the curves were shifted by constant offsets for visual clarity.}
        \label{FigS4}
\end{figure}

To reproduce the SSL state on the $J_1$-$J_2$-$J_3$ triangular lattice, we performed spin dynamics simulations using the Landau-Lifshitz dynamics approach as it implemented in the \textsc{Su(n)ny} program package~\cite{Zhang_2021}. The spins of Cr ions were represented as classical magnetic dipoles, and Heisenberg exchange interactions up to the third-nearest neighbor [as shown in Fig.~1(b) of the main text] were taken into account. The first-nearest neighbor exchange parameter $J_1$ was fixed to be ferromagnetic (FM), and all other exchange parameters are presented in units of $|J_1|$. In calculations, we assumed small easy $ab$-plane anisotropy, consistent with the previous reports~\cite{Baenitz_2021}. We consider a single triangular lattice layer, as the interlayer interaction $J_c$ leads to only a trivial antiferromagnetic (AFM) coupling of the subsequent layers, which, in turn, responsible for the $L = 3/2$~r.l.u. component in the ordering vector $\mathbf{q}_\mathrm{m} = (0.045~0.045~3/2)$~r.l.u.. Moreover, the test calculations on the multilayered system showed that the combination of the FM $J_1$ and the AFM $J_c$ exchanges resulted only in the commensurate AFM magnetic order with the propagation vector $\mathbf{Q} = (0\,0\,3/2)$, indicating that the further-neighbor in-plane AFM exchange interactions are essential for the formation of the spiral order.

To achieve thermal equilibrium for the ground state and the structure factors calculations, a preliminary thermalization was performed. During the annealing process, the temperature was gradually lowered, while the spin configurations were sampled using Langevin dynamics~\cite{Dahlbom_2022}. To balance computational cost and accuracy, a system with periodic boundary conditions and the chosen size of $N\times{N}$ spins with $N~=~300$.

Even though the exact parameters of magnetic interactions in \agcr\ are currently unknown, it is still possible to achieve qualitative description of its magnetic properties. Focusing on the incommensurate phase III with ordering vector $\mathbf{q}_\mathrm{m} = (\xi\,\xi\,0)$, through the classical spin dynamics simulations we derive main general features attributed to \agcr.

\begin{figure}
\includegraphics[width=0.99\linewidth]{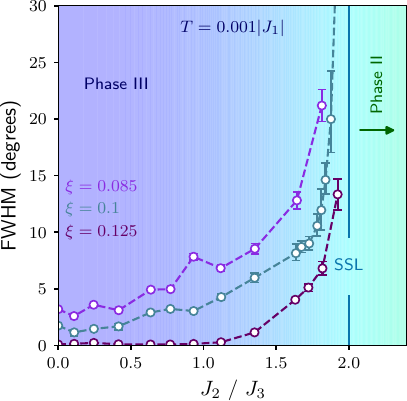}\vspace{3pt}
        \caption{~Dependence of azimuthal width on the $J_2/J_3$ ratio for different magnitudes of the propagation vector.}
        \label{FigS5}
\end{figure}

The magnitude of the incommensurate propagation vector in the ground state depends on the values of $J_2$ and $J_3$ exchange interactions [see the phase diagram in Fig.~1(b) of the main text].
For the phase III, the relation is given by~\cite{Glittum_2021}:
\begin{align}
    \xi = \mathrm{arccos}\Big(\frac{2J'_2-2J'_3-\sqrt{ (3J'_2 + 2J'_3)^2 + 8J'_3 }}{-8J'_3}\Big),
    \label{eq:propagation}
\end{align}
where $J'_2 = J_2/J_1$ and $J'_3 = J_3/J_1$. However, a smaller propagation vector magnitude, i.e. larger wavelength of spin modulation in real space, significantly increases computational expenses. Finite size effects become more prominent as the system contains fewer periods of modulated spins, necessitating a larger system size for better convergence and consequently more thermalization steps. Moreover, achieving thermal equilibrium for long-wave modulated spin textures requires more time steps due to the increased size of a magnetic unit cell. For these reasons, we consider a model system with exchange parameters that correspond to the propagation vector $\mathbf{Q} = (0.1~0.1~0)$, resulting in an approximately two times shorter wavelength of in-plane spin modulation as compared to the one observed in \agcr. This magnitude of the propagation vector imposes constraints on $J_2$ and $J_3$, which can be expressed by the following equation for phase III with $\mathbf{q}_\mathrm{m} = (\xi\,\xi\,0)$ as:
\begin{equation}
    J_3 = \frac{-6\cos{(2\pi\xi)}{J_2} + 3J_2 + 1}{\cos{(2\pi\xi)}[8\cos{(2\pi\xi)} - 4]}. \label{eq:QpIII}
\end{equation}
This constraint sets the ratio $J_2/J_3$ as a free parameter that determines proximity of the system to the II--III critical phase boundary on the phase diagram [see Fig.~1(b) of the main text].

The magnetic order in phase III manifests as six peaks in the structure factor, corresponding to symmetry of the underlying crystal lattice. To analyze spin texture, we extract the azimuthal (in-plane angle) width of these peaks, which depends on the system's proximity to the transition from the phase III to the phase II controlled by the ratio $J_2/J_3$. By integrating the calculated structure factor radially, we obtain the azimuthal dependence of the intensity, which is fitted with a combination of Lorentzian peak functions with the identical full-width-half-maximum. An example of such fit is shown in Fig.~\ref{FigS4}(d). It is important to note that the calculated structure factor is discrete, therefore the quality of the rebinned data improves as larger systems have more Fourier modes present. During the fitting procedure, the peak centers are considered fixed, and the widths of all peaks from different domains are constrained to be equal.

In this type of calculations, a finite system size can naturally affect the simulation results. Therefore, it is essential to ensure that the degree of associated artifacts remains controllable. The system size effects were tested by checking the azimuthal width in relation to the number of spins in the system [Fig.~\ref{FigS4}(c)]. From the tests, we determine that the size less than $200\times200$ spins can introduce finite size effects that lead to enhanced peak width and large uncertainties in its determination. Note that the tests were carried out with exchange parameters corresponding to $\mathbf{q}_\mathrm{m}~=~(0.1\,0.1\,0)$, meaning that, for smaller propagation vectors, the larger system size and relaxation time are necessary in order to sustain this accuracy.

\begin{figure}
\includegraphics[width=0.99\linewidth]{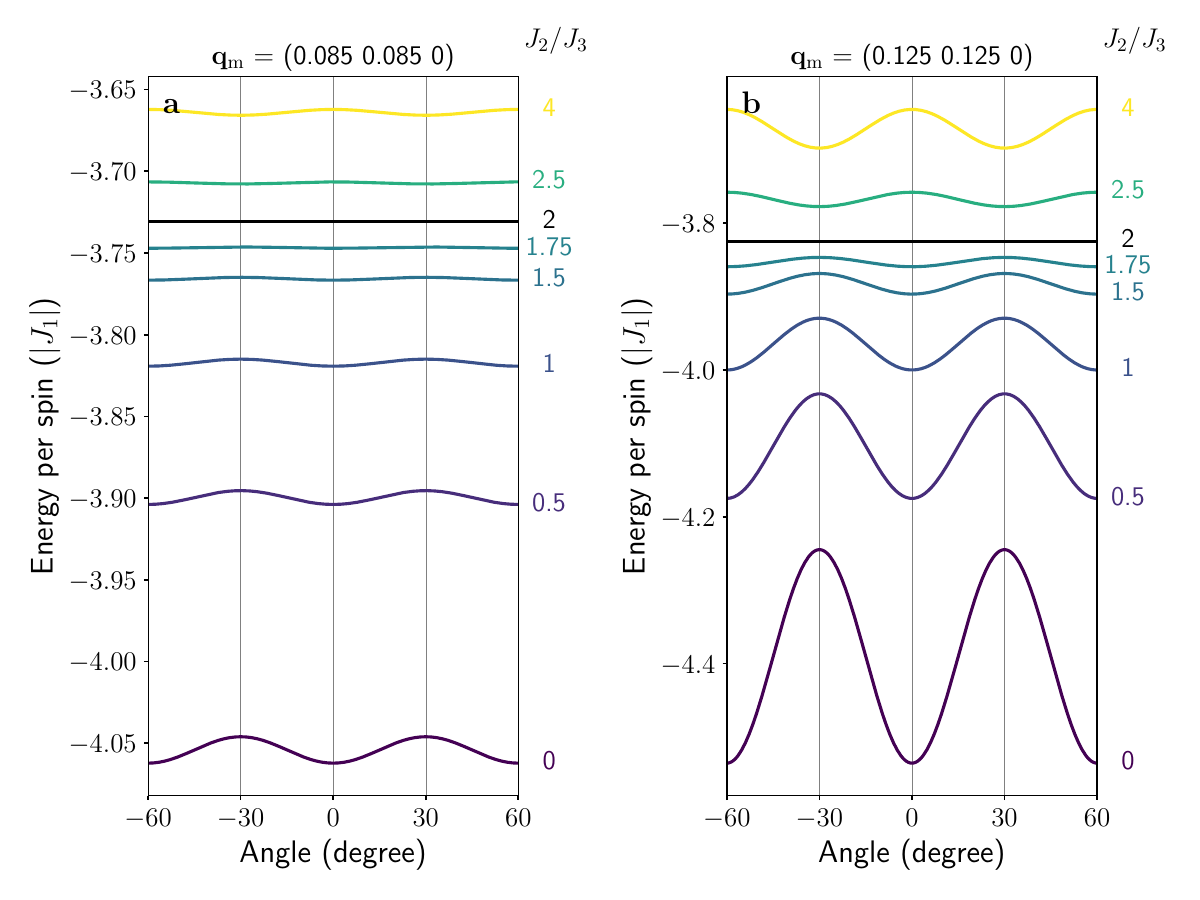}\vspace{3pt}
        \caption{~Analytically calculated energy of a spin spiral as a function of the orientation of its propagation vector. Different curves correspond to different $J_2/J_3$ ratios. Zero angle is chosen as the (110) direction (the phase III) and the absolute value of the propagation $\mathbf{q}_\mathrm{m}~=~(0.085\,0.085\,0)$ for panel (a) and $\mathbf{q}_\mathrm{m}~=~(0.125\,0.125\,0)$ for panel (b).}
        \label{FigS6}
\end{figure}

As mentioned earlier, the azimuthal width gradually increases as the exchange ratio $J_2/J_3$ approaches the value of 2.0, eventually becoming excessively broad. At this region, the width exceeds 60 degrees, compromising the reliability of the Lorentzian fit. This can be clearly seen on Fig.~\ref{FigS5}, where all data points in the figure calculated on the same convergence parameters and system size $300\times300$. The system was thermalized at fixed temperature $k_{\rm B}T~=~0.001|J_1|$ after annealing from $k_{\rm B}T~=~2|J_1|$. Since the chosen convergence parameters were optimized for $\xi~=~0.1$, they become excessive for $\xi~=~0.125$, resulting in decreased uncertainty. Oppositely, this produces a higher level of noise for $\xi~=~0.085$. Nevertheless, it is clear that the azimuthal width is generally higher at a longer wavelength of the spin modulations.

The observed trend of the width response to the propagation vector length plausibly originates from the temperature condition in our simulations. To illustrate this, we analytically calculated the classical Heisenberg energy of a spiral state as a function of the spiral in-plane orientation angle. In Fig.~\ref{FigS6}, the zero angle was chosen along the (110) reciprocal-lattice direction [$\mathbf{q}_\mathrm{m}~=~(0.085\,0.085\,0)$~r.l.u. in panel (a) and $\mathbf{q}_\mathrm{m}~=~(0.125\,0.125\,0)$~r.l.u. in panel (b)]. As it can be seen, the energy has 60~degrees period of the in-plane orientation angle in accord to the underlying crystal-lattice symmetry. Changing frustration via the $J_2/J_3$ ratio, we can transit from the phase III (minimum energy is at $n\cdot30^{\circ}$ for $n$---even) into the phase II (minimum energy is at $n \cdot 30^{\circ}$ for $n$---odd) through the SSL state with maximum frustration level of $J_2/J_3~=~2$, where the energy no longer depends on the propagation vector direction. It is important to note that all the features are preserved when the wavelength of the spiral is modified. However, the energy difference of the 30 and 0 degree spirals (i.e. the energy difference between the phase II and III) also depends on the total energy of the spiral, which, in turn, depends on the modulation wavelength. As was discussed earlier, in the simulations we extract the width of the peaks in the structure factor at a fixed temperature $k_{\rm B}T~=~0.001|J_1|$. In classical dipolar-spin simulations, the azimuthal width of the peaks is directly proportional to the softness of the spin spiral with respect to its in-plane orientation (the ``transverse'' stiffness). This energy dependence surely makes sense only in comparison with the temperature of a system. 
This means that the distribution of spiral orientation is different for a fixed temperature but different spiral period: shorter spiral would have more narrow profile as compared to long-period spiral due to reduced stiffens in the later case.

\section{Simulations of the pancake state at high temperatures}

\begin{figure}
\includegraphics[width=0.99\linewidth]{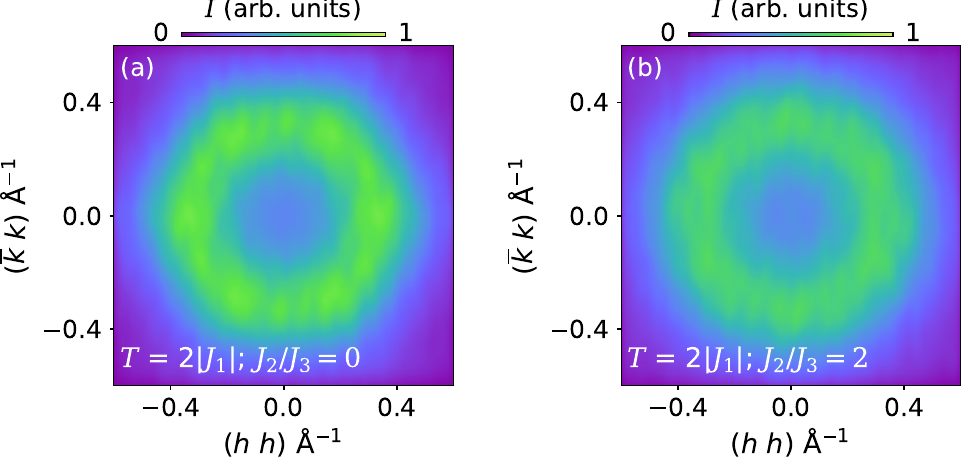}\vspace{3pt}
        \caption{~The calculated structure factor in the paramagnetic state with correlations at finite momentum at $T = 2|J_1|$ for for $J_2/J_3 = 0$ (a)~and $J_2/J_3 = 2$ (b). The convolution with the same experimental resolution was applied as in the main text.}
        \label{Sparamagnet}
\end{figure}

The structure factor of the paramagnetic state with correlations at finite propagation vector is demonstrated in Figs.~\ref{Sparamagnet} where it is shown for the parameters $J_2/J_3 = 0$ (no frustration) and $J_2/J_3 = 2$ (maximal frustration). As can be seen, this high-temperature state is characterized by a broad ring of intensity regardless of the frustration ratio. Unlike the SSL state, which has a well-defined static spin-spiral length, the correlated paramagnet in Figs.~\ref{Sparamagnet} represent spin fluctuations over a wide range of wavelengths.

\begin{figure}
\includegraphics[width=0.99\linewidth]{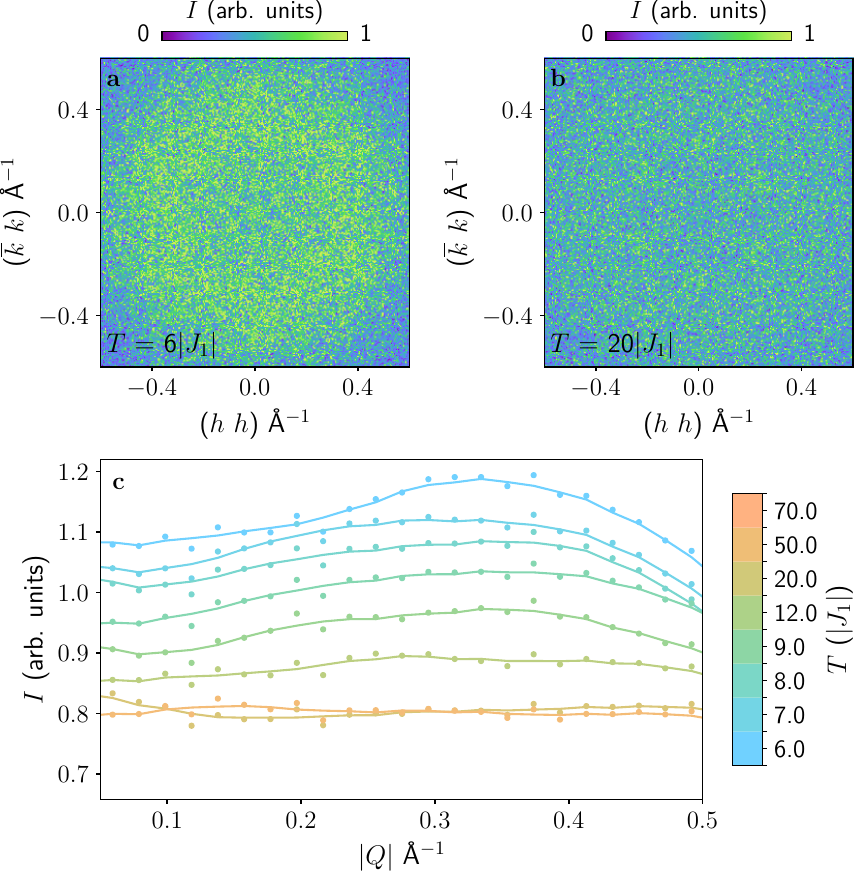}\vspace{3pt}
        \caption{~The calculated structure factor at high temperature. (a,b)~The structure factor at $T = 6|J_1|$ and $20|J_1|$, no convolution with experimental resolution applied. (c)~The radial dependence of the intensity at elevated temperatures.}
        \label{Spancake}
\end{figure}

At even higher temperatures, the broad ring gradually transforms into uncorrelated state~[Fig.~\ref{Spancake}(a,b)]. An intermediate regime with finite residual correlations are referred in literature as ``the pancake state''~\cite{Shimokawa_2019, Yan_2022, Gonzalez_2024}. In our simulations, these correlations are generally weak at $T \gtrsim 3|J_1|$ and diminish to full extent at temperatures above $T \gtrsim 20|J_1|$ [Fig.~\ref{Spancake}(c)]. The same model parameters as in the main text were used for simulations shown in Fig.~\ref{Spancake}: $J_2~= 0.33|J_1|$, $J_3 = 0.19|J_1|$ and $K = 0.03|J_1|$.

\section{High temperature data and background subtraction in the SANS-I experiment} \label{SIbg}

At high temperatures, the SSL state expectedly looses its correlation length. At temperatures much greater than the crossover temperature, the spin spiral pitch is no longer preserved. As it fluctuates around a mean value, the ring-shaped structure factor broadens in the radial direction. Because the SSL continuously transforms into the fully uncorrelated paramagnet over a wide temperature range, this intermediate partially-correlated state received a separate name ``the pancake state'' in the previous studies~\cite{Shimokawa_2019, Yan_2022, Gonzalez_2024}. It is characterized by a broad structure factor, which describes both periodicity and orientation fluctuations of the spin spirals across the system. The structure factor eventually turns into a broad hump with the maxima at zero momentum at sufficiently high temperatures.

\begin{figure}\vspace{8pt}
\includegraphics[width=0.99\linewidth]{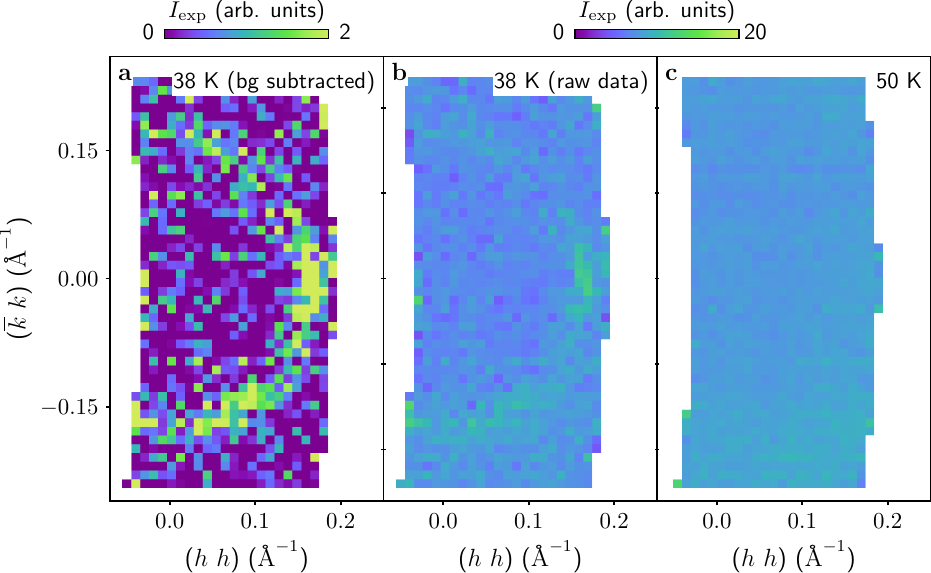}\vspace{3pt}
        \caption{~The background subtraction in the SANS-I experiment. (a)~Diffraction map at 38~K after background subtraction. (b)~The same diffraction map at 38~K but before the background subtraction (raw data). (c)~Diffraction map at 50~K, used as a background.}
        \label{Sparexp}
\end{figure}

Since the intensity pattern of the pancake state is very broadened, it becomes very challenging to observe it in a neutron scattering experiment. The measurements are always accompanied by finite background signal coming from the incoherent scattering or scattering on the sample environment. Therefore, a weak scattering from the expected pancake state appears indistinguishable from the background level. Figure~\ref{Sparexp}(c) shows the raw data collected at 50~K, where the inhomogeneous intensity (higher at the bottom) represent the background contained also in the measurements for all the other temperatures below 50~K. Because the pattern at 50~K is seemingly dominated by the background (the scattering from the pancake state is supposedly lower), we can use for background subtraction for all the measurements shown in the main text. Figures~\ref{Sparexp}(a) and \ref{Sparexp}(b) demonstrate the SSL state at 38~K before and after the background subtraction. It can be seen, that the background-subtracted pattern allows for a more accurate analysis. Further experiments on significantly larger single crystals, which can improve the signal-to-noise ratio, are required for reliable observation of the weakly-correlated pancake state.

\section{Low-temperature defects of the SSL} \label{SIdefects}

\begin{figure*}[t]
\includegraphics[width=0.99\linewidth]{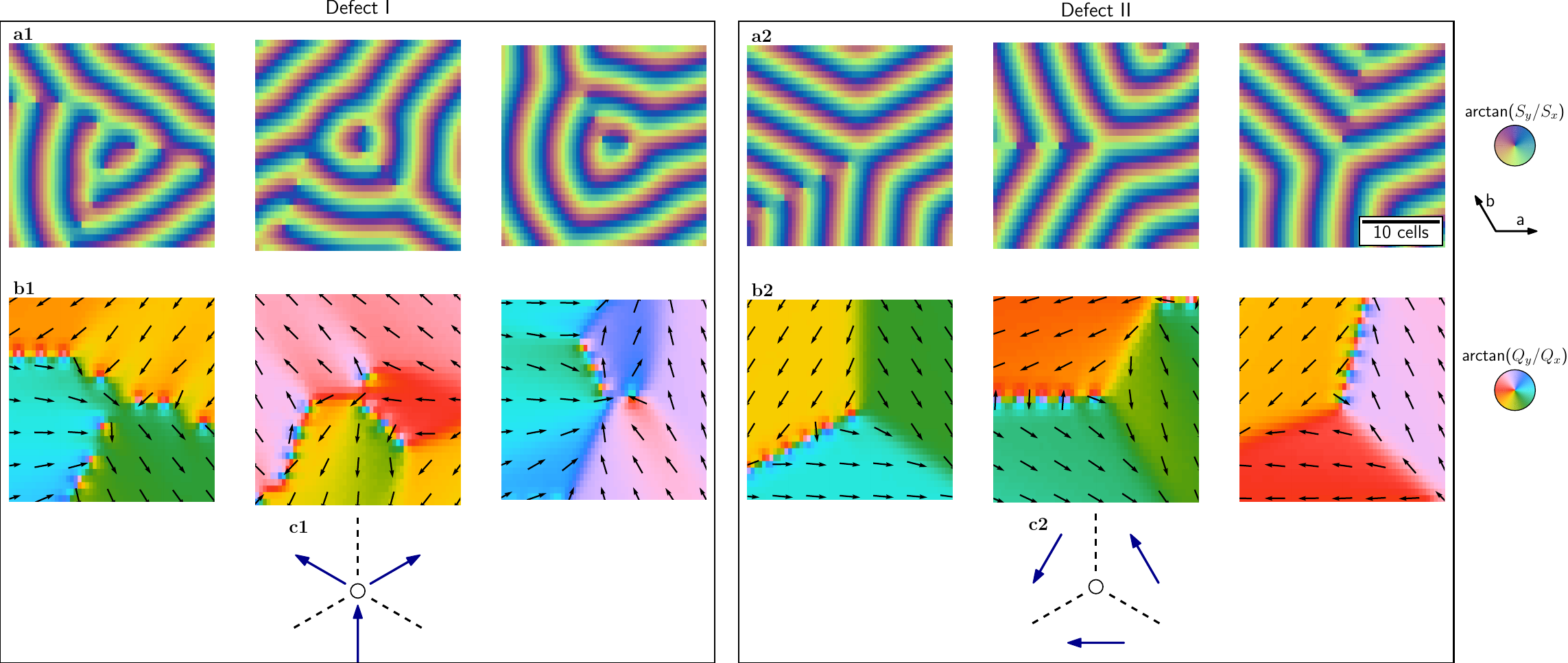}\vspace{3pt}
        \caption{~Examples of the defects found in our simulations. Configurations were obtained after thermalization at $T = 0.0001|J_1|$ and the same model parameters as in the main text were used: $J_2~= 0.33|J_1|$, $J_3 = 0.19|J_1|$ and $K = 0.03|J_1|$. (a1,a2)~Color-coded in-plane direction of the spins in the vicinity of the defects of type I and II. (b1,b2)~The corresponding orientation of the spiral propagation vector in the vicinity of the defects of type I and II. (c1,c2) Schematics of the propagation vector junction of the defects of type I and II.} 
        \label{DefectCalc} 
\end{figure*}

It should be noted, that at low temperatures, when the single ion magnetic anisotropy of \agcr\ well overcomes its temperature fluctuations, the model used in our studies effectively turns into the XY model. 
The XY nature of the spins in \agcr\ at low temperatures thus make it distinguished from the previously reported SSL materials such as FeCl$_3$~\cite{Gao_2022} and MnSc$_2$S$_4$~\cite{Gao_2016, Gao_2020} described by the Heisenberg 3D spins. Therefore, \agcr\ represents a material for which the previous theoretical findings within the XY model of the SSL can be tested.

Yan and Reuther~\cite{Yan_2022} considered the square lattice XY spin model of up to three near-neighbor exchange interactions. Similarly to the triangular-lattice model discussed in the main text, the square model of~\cite{Yan_2022} predicts the SSL state for a certain combination of the frustrating exchange parameters. The Monte Carlo simulations performed by the authors of~\cite{Yan_2022} showed that the stable spin configurations of the SSL may contain features that can be characterized as \textit{momentum vortices}. The latter should not be confused with \textit{the spin vortices}, such as the magnetic skyrmions and the magnetic bubbles, because the momentum vortices, as introduced in~\cite{Yan_2022}, are defined with respect to the vector field of the local directions of the spiral propagation vectors. In a full analogy with the spin vortices, the momentum vortices can have a winding number (the topological charge) attributed to them, which can take values $+1$ for the vortex, $-1$ for the antivortex, and 0 for a nontopological defect~\cite{Yan_2022}. In configurations obtained with our Landau-Lifshitz dynamics calculations at a low temperature, similar momentum defects may occur.

Figures~\ref{DefectCalc}(a1) and \ref{DefectCalc}(a2) demonstrate selected spots within a larger simulated system with two types of the defects found in our model of \agcr, which we label as the Defects~I [Fig.~\ref{DefectCalc}(a1)] and the Defects~II [Fig.~\ref{DefectCalc}(a2)]. To show that all the spin configurations within Fig.~\ref{DefectCalc}(a1) are equivalent but different from the spin textures of Fig.~\ref{DefectCalc}(a2) (which are also mutually equivalent within their set), we turn to the maps of the local propagation vectors where the momentum vortices can be identified. The momentum maps are calculated using equation
\begin{equation}
    Q(\mathbf{r}) = \nabla \arctan(S_y / S_x),
    \label{eq:QQQ}
\end{equation}
where $S_x$ and $S_y$ are the two Cartesian in-plane components of the spin.

As can be seen in Fig.~~\ref{DefectCalc}(b1), each of the momentum map is characterized by a point where three different spiral regions, each with a well defined propagation vector direction $\sim$120~deg apart, join together. The three momenta around the joint form the ``two-in, one-out'' pattern, as schematically depicted in Fig.~\ref{DefectCalc}(c1). Whenever such a momentum junction is stabilized within the SSL state, the spin texture in the vicinity of it is found to feature a ``bubble'' (or a ``loop'') like the ones shown in Fig.~\ref{DefectCalc}(a1).

The spin textures reminiscent of the Y domain walls in Fig.~\ref{DefectCalc}(a2) appear as the result of the momentum vortices shown in Fig.~\ref{DefectCalc}(b2). There, a three momenta encircle the joint point in the fashion of Fig.~\ref{DefectCalc}(c2). As can be seen, the momentum vortex of the Defect~II is rotation invariant to the Defect~I. Namely, all the momenta are equally rotated by 90~deg.

We note that the Defect~II structures are very similar to the configurations of winding number 0 predicted in~\cite{Yan_2022}. Besides, the authors of~\cite{Yan_2022} predicted the topological configurations of winding numbers $+1$ and $-1$, which represent the ``all-in'' (contrary to ``two-in, one-out'') momentum pattern around the joint point. Such defects were not observed in our simulations.

\end{document}